\documentclass{aa} % for a referee version
\usepackage{graphicx}
\usepackage{txfonts}
 \usepackage{lscape}
\begin{document} 
\newcommand{\NB}[2][]{\textbf{\color{red}{#1 #2}}}

  \title{Planet transit and stellar granulation detection with interferometry}
\subtitle{Using the three-dimensional stellar atmosphere \textsc{Stagger}-grid simulations}
\titlerunning{Planet transit and stellar granulation detection with interferometry}
  \author{A. Chiavassa \inst{1}, R. Ligi\inst{1}, Z. Magic\inst{2}, R. Collet\inst{3}, M. Asplund\inst{3}, D. Mourard\inst{1}}
\authorrunning{A. Chiavassa et al.}
\institute{ Laboratoire Lagrange, UMR 7293, CNRS, Observatoire de la C\^ote d'Azur, Universit\'e de Nice Sophia-Antipolis, Nice, France \\
\email{andrea.chiavassa@oca.eu} \and 
Max-Planck-Institut f\"ur Astrophysik, Karl-Schwarzschild-Str. 1, 85741 Garching, Germany  \and 
 Research School of Astronomy $\&$ Astrophysics, Australian National University, Cotter Road, Weston ACT 2611, Australia}
  
   \date{...; ...}

% \abstract{}{}{}{}{} 
% 5 {} token are mandatory
 
  \abstract
   {Stellar activity, and in particular convection-related surface structures, potentially cause bias in the planet detection and characterisation. In the latter, interferometry can help to disentangle the signal of the transiting planet.}
   {We used realistic three-dimensional (3D) radiative hydrodynamical (RHD) simulations from the \textsc{Stagger}-grid and synthetic images computed with the radiative transfer code {{\sc Optim3D}} to provide interferometric observables to extract the signature of stellar granulation and transiting planets.}
   {We computed intensity maps from RHD simulations and produced synthetic stellar disk images as a nearby observer would see accounting for the centre-to-limb variations. We did this for twelve interferometric instruments covering wavelengths ranging from optical to infrared. We chose an arbitrary date and arbitrary star with coordinates that ensures observability throughout the night. This optimization of observability allows for a broad coverage of spatial frequencies. The stellar surface asymmetries in the brightness distribution, either due to convection-related structures or a faint companion mostly affect closure phases. We then computed closure phases for all images and compared the system star with a transiting planet and the star alone.
We considered the impact of magnetic spots constructing a hypothetical starspots image and compared the resulting closure phases with the system star with a transiting planet.}
   {We analyzed the impact of convection at different wavelengths. All the simulations show departure from the axisymmetric case (closure phases not equal to 0 or $\pm\pi$) at all wavelengths. The levels of asymmetry and inhomogeneity of stellar disk images reach high values with stronger effects from 3rd visibility lobe on. We presented two possible targets  (Beta Com and Procyon) either in the visible and in the infrared and found that departures up to 16$^\circ$ can be detected on the 3rd lobe and higher. In particular, MIRC is the most appropriate instrument because it combines good UV coverage and long baselines. Moreover, we explored the impact of convection on interferometric planet signature for three prototypes of planets with sizes corresponding to one hot Jupiter, one hot Neptune, and a terrestrial planet. The signature of the transiting planet on closure phase is mixed with the signal due to the convection-related surface structures, but it is possible to disentangle it at particular wavelengths (either in the infrared or in the optical) by comparing the closure phases of the star at difference phases of the planetary transit. It must be noted that starspots caused by the magnetic field may pollute the granulation and the transiting planet signals. However, it is possible to differentiate the transiting planet signal because the time-scale of a planet crossing the stellar disk is much smaller than the typical rotational modulation of a star.}
   {The detection and characterisation of planets must be based on a comprehensive knowledge of the host star; this includes the detailed study of the stellar surface convection with interferometric techniques. In this context, RHD simulations are crucial to reach this aim. We emphasize that interferometric observations should be pushed at high spatial frequencies by accumulating observations on closure phases at short and long baselines. }


   \maketitle
%
%________________________________________________________________

\section{Introduction}

 \begin{table*}
\centering
\begin{minipage}[t]{\textwidth}
\caption{3D simulations from \textsc{Stagger}-grid used in this work.}             % title of Table
\label{simus}      % is used to refer this table in the text
\centering                          % used for centreing table
\renewcommand{\footnoterule}{} 
\begin{tabular}{c c c c c c c c}        % centreed columns (4 columns)
\hline\hline                 % inserts double horizontal lines
$<T_{\rm{eff}}>$\footnote{Horizontally and temporal average of the emergent effective temperatures from \cite{2013A&A...557A..26M}} & [Fe/H]  & $\log g$ & $x,y,z$-dimensions & $x,y,z$-resolution   & $\rm{M}_{\star}$ & $\rm{R}_{\star}$ & Number of tiles  \\
$[\rm{K}]$ & & [cgs]  & [Mm]  & [grid points]   & [$\rm{M}_\odot$] & [$\rm{R}_\odot$] & over the diameter\\
\hline
5768.51 (Sun) & 0.0 & 4.4 &  3.33$\times$3.33$\times$2.16 & 240$\times$240$\times$240 & 1.0 & 1.0 & 286\\
5764.13 & -1.0 & 4.4 &  3.12$\times$3.12$\times$1.63  & 240$\times$240$\times$240 & 1.0 & 1.0 & 305\\
5781.04 & -2.0 & 4.4 &  2.75$\times$2.75$\times$1.67  & 240$\times$240$\times$240 & 1.0 & 1.0 & 347\\
5780.06 & -3.0 & 4.4 &  3.00$\times$3.00$\times$1.61  & 240$\times$240$\times$240 & 1.0 & 1.0 & 318\\
4569.23 & 0.0 & 2.0 &  1000$\times$1000$\times$1288 & 240$\times$240$\times$240 & 1.3 \footnote{Averaged value from Fig.~4 of \cite{2012A&A...537A..30M}} & 18.9 & 17\\
5001.35 & 0.0 & 3.5 &  27.08$\times$27.08$\times$24.49  & 240$\times$240$\times$240 & 1.15 \footnote{Averaged value from Fig.~2 of \cite{2011ApJ...740L...2S}} & 3.1 & 121\\
5993.42 & 0.0 & 4.0 &  10.83$\times$10.83$\times$5.66 & 240$\times$240$\times$240 & 1.0 $^{c}$ & 1.6 & 266\\
5998.93 & 0.0 & 4.5 &  2.92$\times$2.92$\times$1.76  & 240$\times$240$\times$240 & 1.15 $^{c}$& 0.99 & 312\\

\hline\hline                          % inserts single horizontal line
\end{tabular}
\end{minipage}
\end{table*}

Two very successful methods for finding exoplanets orbiting around stars are the transiting and radial velocity methods. The transit happens when a planet passes between the exoplanet and its host star. The planet then blocks some of the star-light during the transit and creates a periodic dip in the brightness of the star. Observations taken during both the primary and secondary transit can be used to deduce the composition of the planet's atmosphere. \\
As the star moves in the small orbit resulting from the pull of the exoplanet, it will move towards the planet and then away as it completes an orbit. Regular periodic changes in the star's radial velocity (i.e., the velocity of the star along the line of sight of an observer on Earth) depend on the planet's mass and the inclination of its orbit to our line of sight. Measurements on the Doppler-shifted spectra give a minimum value for the mass of the planet.

However, a potential complication to planet detection may be posed by stellar surface inhomogeneities (due to the presence of stellar granulation, magnetic spots, dust, etc.) of the host star. In this article we investigate in particular problem of stellar granulation. It was first observed on the Sun by \cite{1801RSPT...91..265H} and today modern telescopes provide direct observations \citep[e.g., ][]{2004ApJ...610L.137C}. However, the best observational evidence comes from unresolved spectral line in terms of widths, shapes, and strengths that, when combined with numerical models of convection, allow quite robust results to be extracted from the simulations \citep{2009LRSP....6....2N,2000A&A...359..669A}. For this purpose, large efforts have been made in recent decades to use theoretical modeling of stellar atmospheres to solve multidimensional radiative hydrodynamic equations in which convection emerges naturally. These simulations take into account surface inhomogeneities (i.e., granulation pattern) and velocity fields. The widths of spectral lines are heavily influenced by the amplitude of the convective velocity field, which overshoots into the stable layers of the photosphere where the lines are formed. This results in characteristic asymmetries of spectral lines as well as net blueshifts \citep[e.g.][]{dravins87}. The observation and interpretation of unresolved stellar granulation is not limited to the Sun \citep{2009LRSP....6....2N} because numerical simulations cover a substantial portion of the Hertzsprung-Russell diagram \citep{2013A&A...557A..26M,2013ApJ...769...18T,2009MmSAI..80..711L}, including the evolutionary phases from the main-sequence over the turnoff up to the red-giant branch for low-mass stars.

Since the discovery of  51 Peg \citep{1995Natur.378..355M}, various studies have looked at starspots. For instance, \cite{1998ApJ...498L.153S} proposed the first quantitative impact of starspots on radial-velocity measurements. The authors studied the impact of these surface structures on the bisector (i.e., measure of the spectral line asymmetries) global slope and found that convection leads to bisector variations up to a few tens ms$^{-1}$. \cite{1997ApJ...485..319S} pointed out they can lead to even larger radial-velocity variations for G2V-type stars. It should be expected that, in the case of F dwarfs or K giants, the velocity fields would be even larger. \cite{2004AJ....127.3579P} measured star-to-star variations of 50 $ms^{-1}$ due to stellar activity in a sample of Hyades dwarfs. \cite{2007A&A...473..983D} discussed the possibility that, in F-K type stars, radial-velocity variations may be due to either spots or planets. \cite{2011ApJ...743...61S} showed that the transit data of a super-Neptune planet exhibit numerous anomalies that they interpret as passages over dark spots.

%Light curve analysis rely on theoretical predictions for limb darkening carried out using classical one dimensional hydrostatic stellar model atmospheres, for which extensive grids are available \citep[e.g., ][]{2000A&A...363.1081C}. However, one dimensional model use an incomplete treatment of convection, the mixing-length theory, and cannot reproduce stellar inhomogeneities. 

The role of long-baseline interferometric observations in planet hunting is a complement to the radial velocity and adaptive optics surveys. Thanks to the higher angular resolution, interferometry is the ideal tool for exploring separations in the range 1 to 50 mas \citep{2012A&A...541A..89L}. This is achieved by observing the closure phase measurements directly associated with the asymmetries in the brightness distribution, and, as a consequence, off-axis detection of a companion. Long-baseline interferometry bridges the gap between the use of direct imaging, which finds wide companions, and the use of RV measurements, which detect close companions \citep{2012A&A...541A..89L}. Several attempts and discussions regarding prospective ideas towards this end have already been carried out. In particular for hot Jupiter planets, with the MIRC instrument at CHARA telescope \cite[e.g., ][]{2008SPIE.7013E..45Z,2008PASP..120..617V,2011PASP..123..964Z} or the AMBER, MIDI, PIONIER instruments at VLTI \citep[e.g., ][]{2010A&A...515A..69M,2011A&A...535A..68A,2012A&A...540A...5C,2013MNRAS.435.2501L}. However, the extraction of the planetary signal from the interferometric observables is a difficult task that requires very accurate precision levels, possible only with proportionate increase of the data signal to noise.

In this work, we present interferometric predictions obtained from three-dimensional surface convection simulations run for stars spanning different effective temperatures, surface gravities, and metallicities. Further, we present results from a study of the impact of granulation on the detection of transiting planet for three prototypes of planets of different sizes corresponding to a hot Jupiter, a hot Neptune, and a terrestrial planet.

    \begin{figure*}
   \centering
   \begin{tabular}{c}
        \includegraphics[width=0.99\hsize]{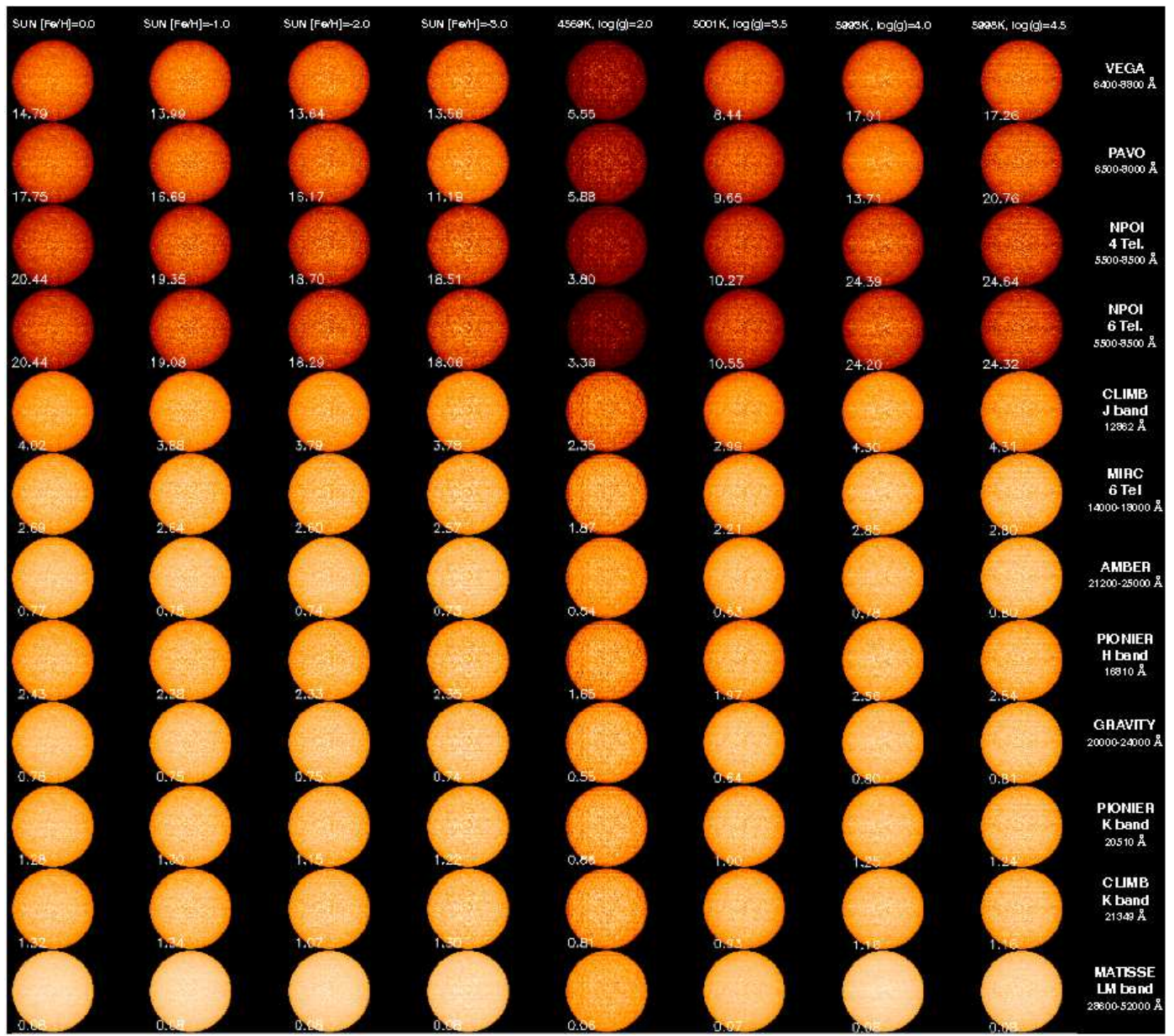}\\
          \end{tabular}
      \caption{Synthetic stellar disk images of the RHD simulations of Table~\ref{simus} (columns). The images correspond to a representative wavelength for each interferometric instruments of Table~\ref{instruments} from the optical (top row) to the far infrared (bottom row). The averaged intensity ($\times10^5$\,erg\,cm$^{-2}$\,s$^{-1}$\,{\AA}$^{-1}$) is reported in the lower left corner of each image.}
        \label{images}
   \end{figure*}
   
   \section{Stellar model atmospheres}

  \cite{2013A&A...557A..26M} described the large \textsc{Stagger}-grid of realistic three-dimensional radiative hydrodynamical (RHD) simulations of stellar convection for cool stars using \textsc{Stagger}-code (originally developed by Nordlund $\&$ Galsgaard 1995\footnote{http://www.astro.ku.dk/$\sim$kg/Papers/MHD\_code.ps.gz}, and continuously improved over the years by its user community), a state-of-the-art (magneto)hydrodynamic code that solves the time-dependent equations for conservation of mass, momentum and energy. The code uses periodic boundary conditions horizontally and open boundaries vertically. At the bottom of the simulation, the inflows have constant entropy and pressure. The outflows are not constrained and are free to pass through the boundary. The code is based on a sixth-order explicit finite-difference scheme, and a fifth-order interpolation. The considered large number over wavelength points is merged into 12 opacity bins \citep{1982A&A...107....1N,2000ApJ...536..465S}. The equation-of-state accounts for ionization, recombination, and dissociation \citep{MHD}. The opacities include continuous absorption and scattering coefficients as listed in \cite{2010A&A...517A..49H}, and the line opacities as described in \cite{2008A&A...486..951G}, in turn based on the VALD-2 database \citep{2001ASPC..223..878S} of atomic lines and the SCAN-base \citep{1997IAUS..178..441J} of molecular lines. \\
  For the solar abundances the authors employed the latest chemical composition by \cite{asplund09}, which is based on a solar simulation performed with the same code and atomic physics as in \cite{2013A&A...557A..26M}.

    \section{Three-dimensional radiative transfer}

    We used pure-LTE radiative transfer \textsc{Optim3D} \citep{2009A&A...506.1351C} to compute synthetic images from the snapshots of the RHD simulations of the \textsc{Stagger}-grid \citep[see Fig.~1 of][]{2013A&A...557A..26M}. The code takes into account the Doppler shifts occurring due to convective motions. The radiative transfer equation is solved monochromatically using pre-tabulated extinction coefficients as a function of temperature, density, and wavelength. \\
    The lookup tables were computed for the same chemical compositions \citep{asplund09} as the RHD simulations using the same extensive atomic and molecular continuum and line opacity data as the latest generation of MARCS models \citep{2008A&A...486..951G}.  We assume zero microturbulence and model the non-thermal Doppler broadening of spectral lines  using only the self-consistent velocity fields issued from the 3D simulations. The temperature and density ranges spanned by the tables are optimized for the values encountered in the RHD simulations. The detailed methods used in the code are explained in \cite{2009A&A...506.1351C}. \textsc{Optim3D} has already been employed in synergy with \textsc{Stagger}-code within several works \citep{2010A&A...524A..93C,2011JPhCS.328a2012C,2012A&A...540A...5C} either concerning the extraction of interferometric observables or synthetic spectra.

 \section{Interferometric observable construction}

  \begin{figure}
   \centering
   \begin{tabular}{c}
              \includegraphics[width=0.9\hsize]{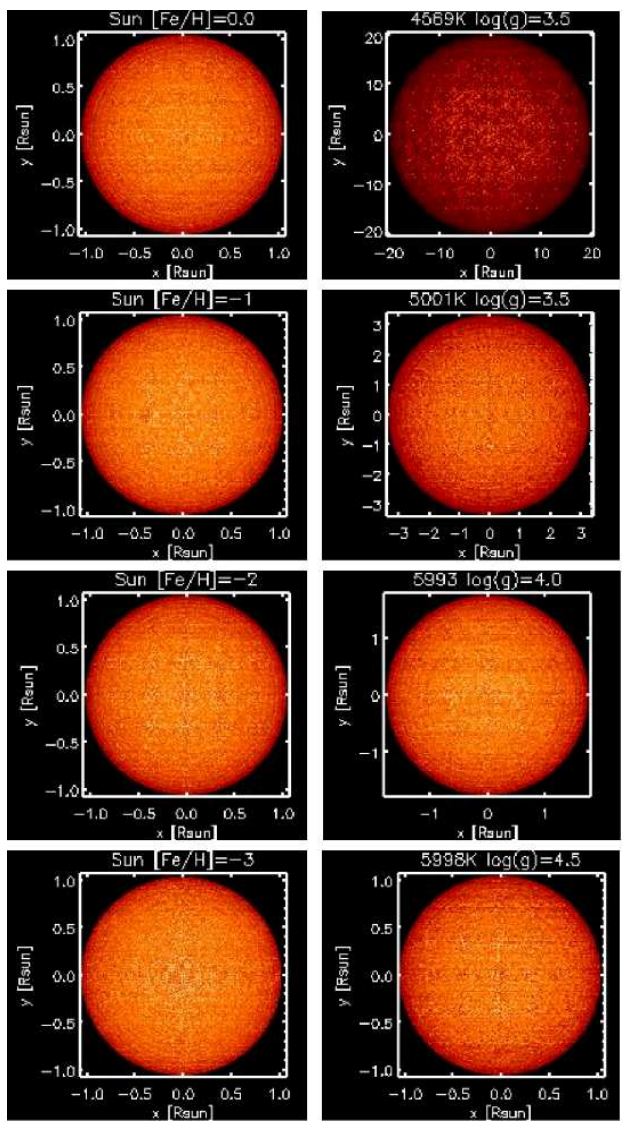}
        \end{tabular}
      \caption{Enlargement of the synthetic stellar disk images of Fig.~\ref{images} for the VEGA instrument (Table~\ref{instruments}).}
              \label{imagesbis}
   \end{figure}

\begin{figure}
   \centering
   \begin{tabular}{c}
        \includegraphics[width=0.98\hsize]{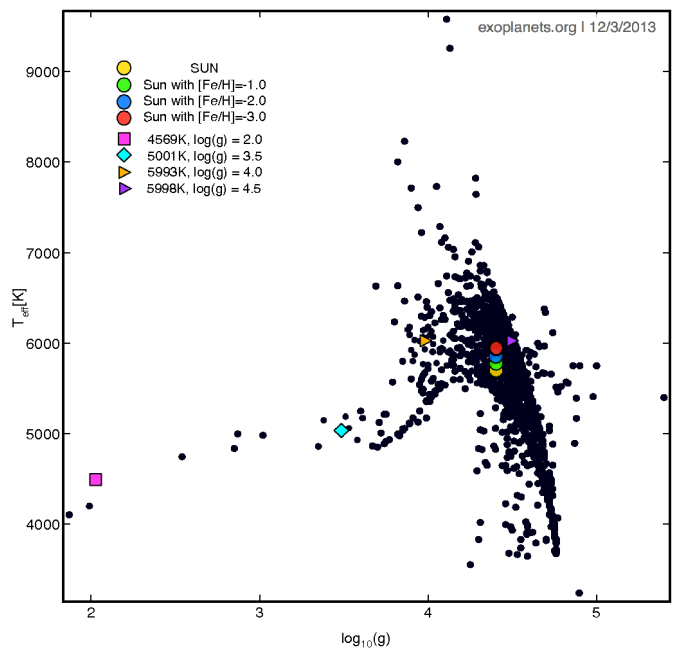}\\
          \end{tabular}
      \caption{RHD simulations from \textsc{Stagger}-grid used in this work with the aimed effective temperature (see also Table~\ref{simus}) over-plotted to the Kepler planets (confirmed and candidates) in fall 2013 from http://exoplanets.org \citep{2011PASP..123..412W}.}
        \label{covering}
   \end{figure}

The aim of the present work is to present a survey of the convective pattern ranging from the optical to the far infrared and to evaluate its effect on the detection of planet transit. We chose representative simulations in the \textsc{Stagger}-grid partially covering the Kepler planets and to study the effect of the metallicity  across the HR-diagram to cover typical Kepler planets, and including different metallicities for the solar model (Fig.~\ref{covering}). Our statistical approach aim to present results that can be extrapolated to other stars in the Hertzsprung-Russel. More detailed analysis with respect to particular stellar parameters can be conducted using specific simulations of the \textsc{Stagger}-grid.

We used \textsc{Optim3D} to compute intensity maps from the snapshots of the RHD simulations of Table~\ref{simus} for different inclinations with respect to the vertical, $\mu{\equiv}\cos(\theta)$=[1.000, 0.989, 0.978, 0.946, 0.913, 0.861, 0.809, 0.739, 0.669, 0.584, 0.500, 0.404, 0.309, 0.206, 0.104] \citep[these angles have already been used in the previous works of ][]{2012A&A...540A...5C,2010A&A...524A..93C}, and for a representative series of 10 snapshots spaced adequately apart so as to capture several convective turnovers for each simulation. The wavelength range is between 4000 and 52000 \AA\ with a spectral resolution $\lambda/\Delta\lambda=20000$.

\subsection{From a small portion of the stellar surface to spherical tile images}\label{tilingsect}

   \begin{figure}
   \centering
   \begin{tabular}{c}
              \includegraphics[width=0.92\hsize]{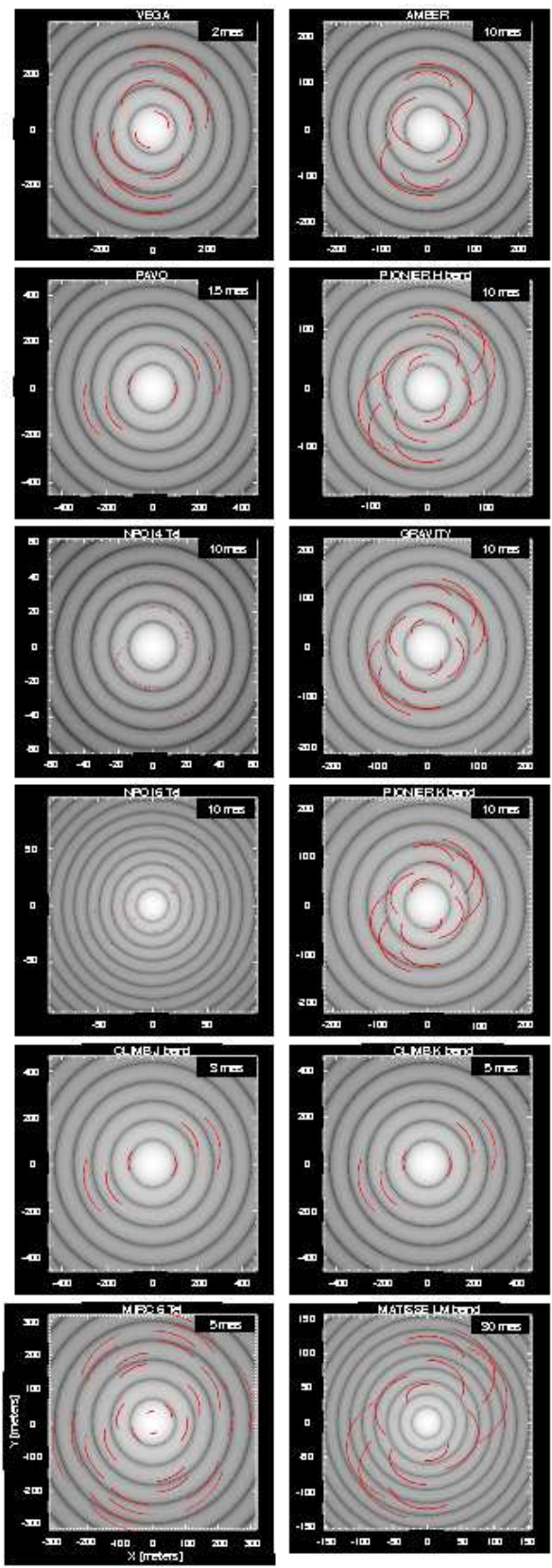}
        \end{tabular}
              \caption{Typical UV coverage in meters for the different instruments of Table~\ref{instruments} over-plotted to the Fourier Transform of the RHD simulation of the Sun (Table~\ref{simus}). Red dots correspond to the telescope baseline positions during the arbitrary observation we prepared. The observability is assured for approximatively a whole night with a large enough number of Earth rotation aperture synthesis baseline points. The arbitrary apparent size of the star is reported in the top-right corner of each panel.}
        \label{uvplan}
   \end{figure}

The computational domain of the RHD simulations is limited to a small representative volume located in the stellar photosphere including the top of the stellar convective envelope, the horizontal directions chosen so as to be large enough to cover an area corresponding to about ten granular cells. The intensity maps computed with \textsc{Optim3D} are limited to a small portion of the stellar surface \citep[see, e.g., Fig~1 of ][]{2010A&A...524A..93C}, thus to overcome this limitation, we applied the same method as explained in \cite{2010A&A...524A..93C} to tile a spherical surface accounting for limb-darkened effects. The computed value of the $\theta$-angle used to generate each map depended on the position (longitude and latitude) of the tile on the sphere and was linearly interpolated among the inclination angles.\\
In addition to this, the statistical tile-to-tile fluctuations (i.e., number of granules, shape, and size) is taken in consideration by selecting random snapshots within each simulation's time-series. As a consequence, the simulation assumption of periodic boundary conditions resulted in a tiled spherical surface globally displaying an artifactual periodic granulation pattern. However, \cite{2010A&A...524A..93C} proved that the signal artificially introduced into the interferometric observables is weaker than the signal caused by the inhomogeneities of the stellar surface.\\
We estimated a stellar radius based on an applied mass taken from the literature (Table~\ref{simus}, 6th column), then we computed the number ($N_{\rm{tile}}$) of tiles needed to cover half a circumference from side to side on the sphere $N_{\rm{tile}} = \frac{\pi \cdot \rm{R}_\odot}{x,y\rm{-dimension}}$, where $\rm{R}_\odot$ (transformed in Mm) and  $x,y\rm{-dimension}$ come from Table~\ref{simus}. 
    
 The final result is an orthographic projection of the tiled spheres (Fig.~\ref{images}). It must be noted that our method of constructing realisations of stellar disk images inevitably introduces some discontinuities between neighboring tiles by randomly selecting temporal snapshots and by cutting intensity maps at high latitudes and longitudes. The figure shows that the centre-to-limb variations are more pronounced in the optical instruments with respect to the infrared ones. This effect was already found in \cite{2012A&A...540A...5C,2010A&A...524A..93C} for some K-giant and sub-giant stars and is explained by different sensitivity of the source (Planck) function at optical and at infrared wavelengths.

\subsection{Choice of interferometric instruments}\label{interfere}
 
Actual interferometers ensure the wavelength coverage from optical to far infrared with a series of instruments mounted on different sites. Table~\ref{instruments} displays the instruments we chose, where they are mounted, and the number of telescopes recombined as well as the wavelength probed. 
We used the online Astronomical Software for Preparing Observations (ASPRO2) of the JMMC\footnote{www.jmmc.fr/aspro$\_$page.htm} to extract an OIFITS file with the telescope real positions in the UV-plane, telescope configurations, and observing wavelength. Afterwards, we performed a top-hat average over the whole set of disk images to obtain one synthetic image for each observing wavelength. Even if a wavelength dependence exists on the interferometric observables, for simplicity, we assume a representative wavelength for each instrument in the rest of the work. \\
In Section~\ref{closuresect}, we introduce the closure phase observable and study its potentiality for the detection of surface related convective structures.  We do not aim to interpret/observe a particular star and thus, for each instrument, we chose an arbitrary date and arbitrary star with coordinates that ensure observability for the whole night. This choice is taken to accommodate a large coverage of the spatial frequencies up to the 5th-6th lobe when possible (Fig.~\ref{uvplan}). Due to the sparse selections of baselines (i.e., different apparent size of the targets), using this approach it is not possible to directly compare directly the instruments, however, in this section, we aim to present a closure phase survey of the convective pattern from the optical to the far infrared.\\
Finally, in Section~\ref{realsection}, we investigate a more concrete scenario with the choice of two real targets either in the visible and in the infrared. Thanks to the fact that we consider fixed targets for visible and infrared instruments, we can directly compare the results among the different instruments and propose the best instrument and/or interferometric facility to detect the stellar granulation.
  
\section{Closure phase as an indicator of the stellar inhomogeneity}\label{closuresect}

    \begin{table*}
%     \scriptsize
\tiny
%\footnotesize
\centering
\begin{minipage}[t]{\textwidth}
\caption{Interferometric instruments/configurations used in this work. Some of the instruments may cover other wavelength range or be used with different configurations, but what we chose is a good representation for the purpose of the present work. All the information about the instrument/configuration/wavelengths have been retrieved on ASPRO2, except for VISION (V. E. Garcia, private communication).}             % title of Table
\label{instruments}      % is used to refer this table in the text
\centering                          % used for centreing table
\renewcommand{\footnoterule}{} 
\begin{tabular}{c c c c c c c c c}        % centreed columns (4 columns)
\hline\hline                 % inserts double horizontal lines
Name  & Location & Max  & wavelength  &    Closure  &  Configuration  &  Number of  & Active   & Reference \\
	  &  		     & baseline [m] & range [\AA] &  Phase error [$^\circ$] &  chosen  & telescopes   & since/from & \\
\hline
VEGA & CHARA\tablefootmark{(a)} & $\sim$331 & 	$\sim$6400-8800		  &	---   &  W2-E2-S1-E1 & 4 & 	2009  & 1   \\
PAVO & CHARA & $\sim$331 & $\sim$6500-8000   & --- &  S1-W1-W2 & 3 & 2011 & 2 \\
VISION    & NPOI\tablefootmark{(b)} &  $\sim$432 & $\sim$5500-8500 & --- & AC0-AW0-AN0-E06  & 4 & 2014?& 3  \\
                 &                                           &                     & 				      &      & AC0-AE0-AW0-AN0-E06-W07 &  6 & & 3 \\
CLIMB  & CHARA & $\sim$331 & 12862/21349  & --- & S1-W1-W2 & 3 & 2005 & 4\\
	   &  	 	     &    		 & broad-band   &   &  &  & & \\
MIRC &  CHARA &  $\sim$331   & $\sim$14000-18000  &  0.1-0.2\tablefootmark{(c)} &  S1-S2-W1-W2-E1-E2  & 6 & 2007& 5 \\
AMBER & VLTI \tablefootmark{(d)} & 130 &  $\sim$21200-25000 & 0.20-0.37\tablefootmark{(e)}& A1-G1-J3 & 3  & 2004 &  6 \\
PIONIER & VLTI &  130 & 16810/20510 & 0.25-1\tablefoottext{f} & A1-G1-K0-J3 & 4 & 2010 & 7 \\
	   &  	 	     &    		 & broad-band   &   &  &  & & \\
GRAVITY & VLTI &  130 & $\sim$20000-24000 & 1\tablefoottext{g} & A1-G1-K0-J3 & 4 & 2015? & 8\\
MATISSE & VLTI &  130 & $\sim$28600-52000 & < 1.16\tablefoottext{h} & A1-G1-K0-J3 & 4 & 2016? & 9 \\
\hline\hline                          % inserts single horizontal line
\end{tabular}
\tablefoottext{a}{                     \cite{2005ApJ...628..453T}}
\tablefoottext{b}{\cite{2014AAS...22320202A}}%\cite{1998ApJ...496..550A}}
\tablefoottext{c}{\cite{2011PASP..123..964Z,2010SPIE.7734E..37Z, 2008SPIE.7013E..45Z}}
\tablefoottext{d}{\cite{2008SPIE.7013E..11H}}
\tablefoottext{e}{\cite{2010A&A...520L...2A}  for medium resolution} 
\tablefoottext{f}{\cite{2011A&A...535A..67L, 2011A&A...535A..68A}} 
\tablefoottext{g}{Final Design Review 2011, private communication}
\tablefoottext{h}{\cite{2012POBeo..91..129L}}
\tablebib{(1)~\cite{2009A&A...508.1073M};
(2) \cite{2008SPIE.7013E..63I}; (3) \cite{2012AAS...21944613G}; (4) \cite{2005ApJ...628..453T};
(5) \cite{2004SPIE.5491.1370M}; (6) \cite{2007A&A...464....1P}; (7) \cite{2011A&A...535A..67L};
(8) \cite{2008SPIE.7013E..69E}; (9) \cite{2008SPIE.7013E..70L}.}

\end{minipage}
\end{table*}

\begin{figure*}
   \centering
   \begin{tabular}{cc}
              \includegraphics[width=0.32\hsize]{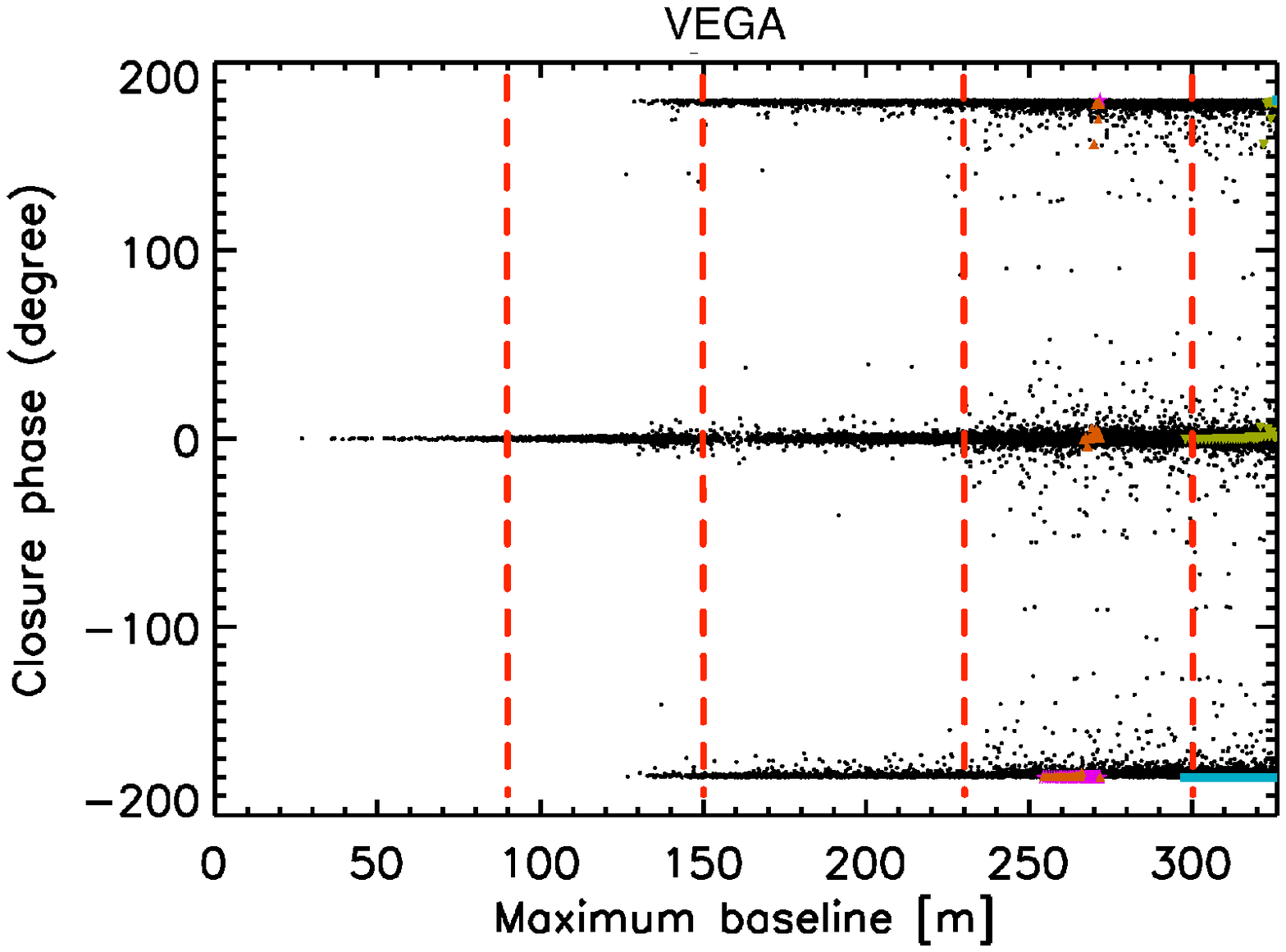}
              \includegraphics[width=0.32\hsize]{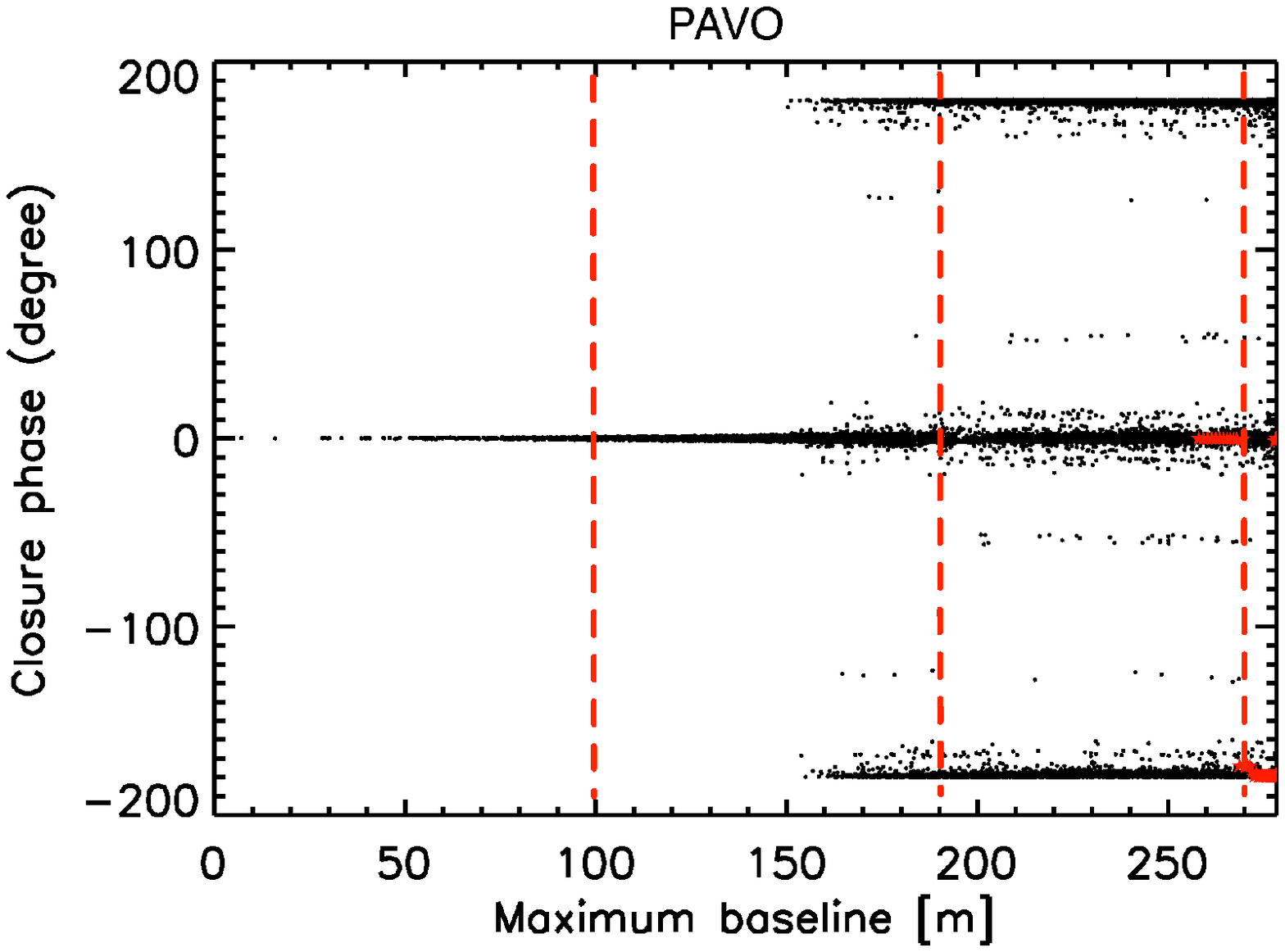}
               \includegraphics[width=0.32\hsize]{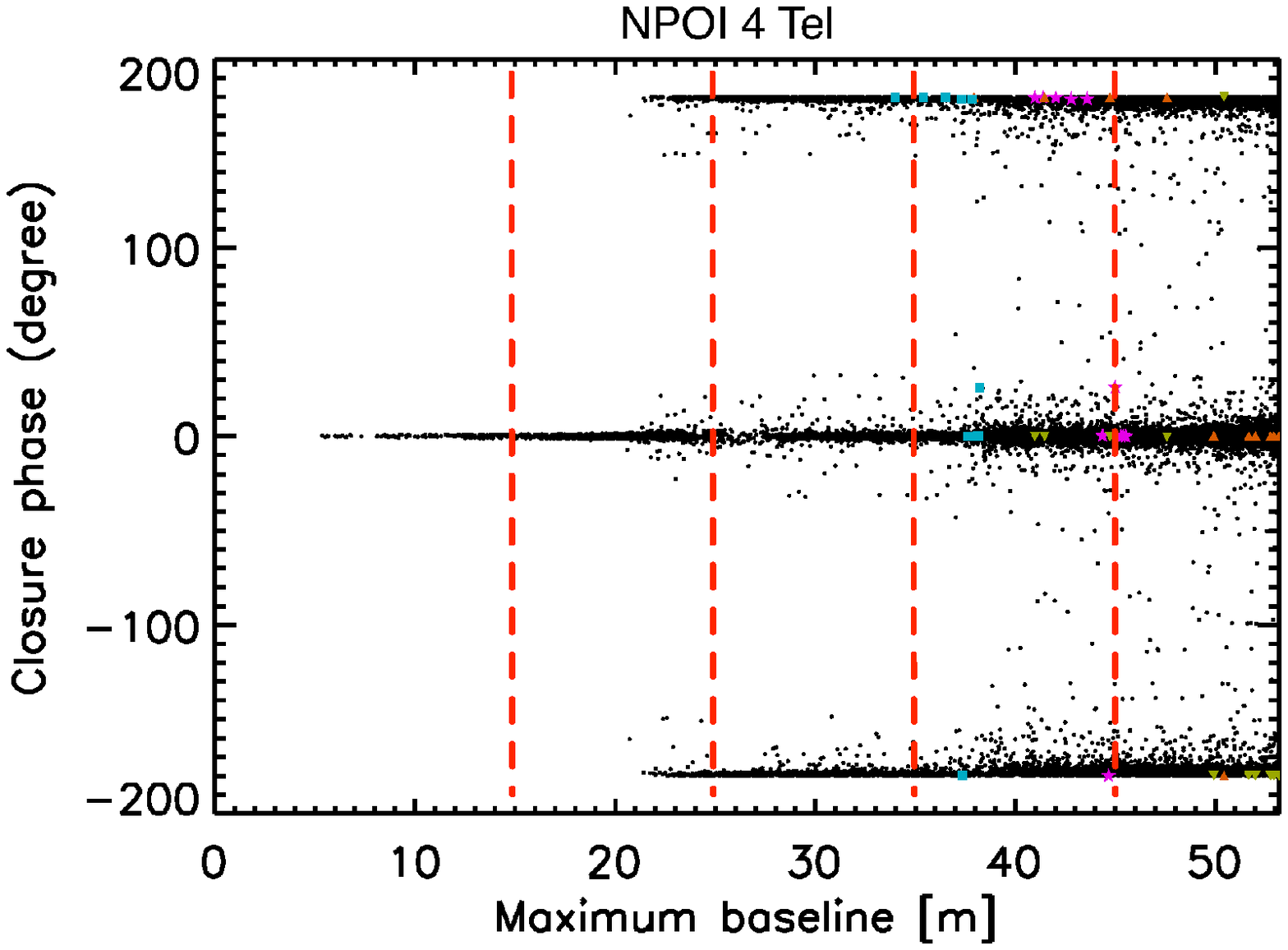}\\
              \includegraphics[width=0.32\hsize]{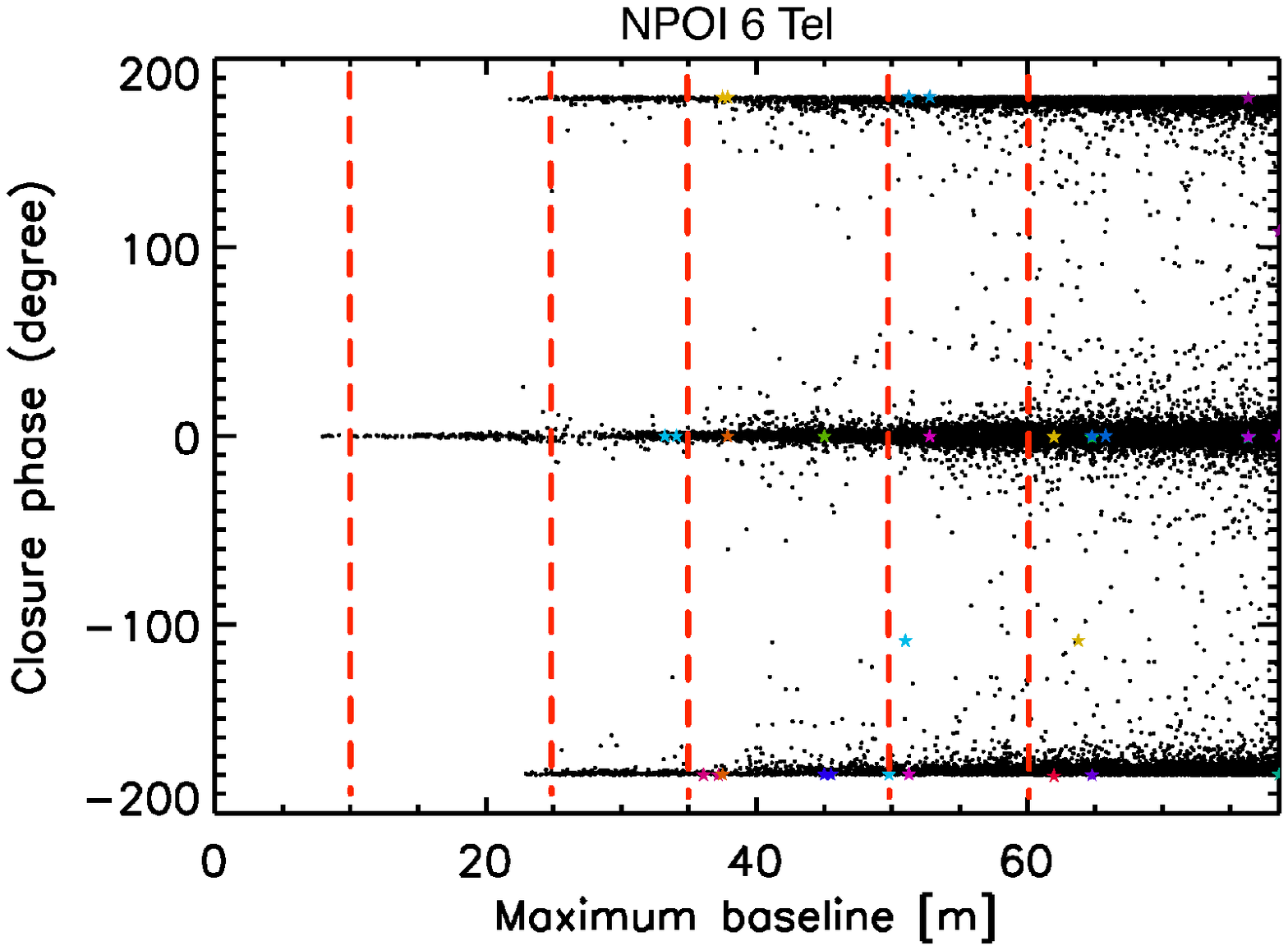}
	   \includegraphics[width=0.32\hsize]{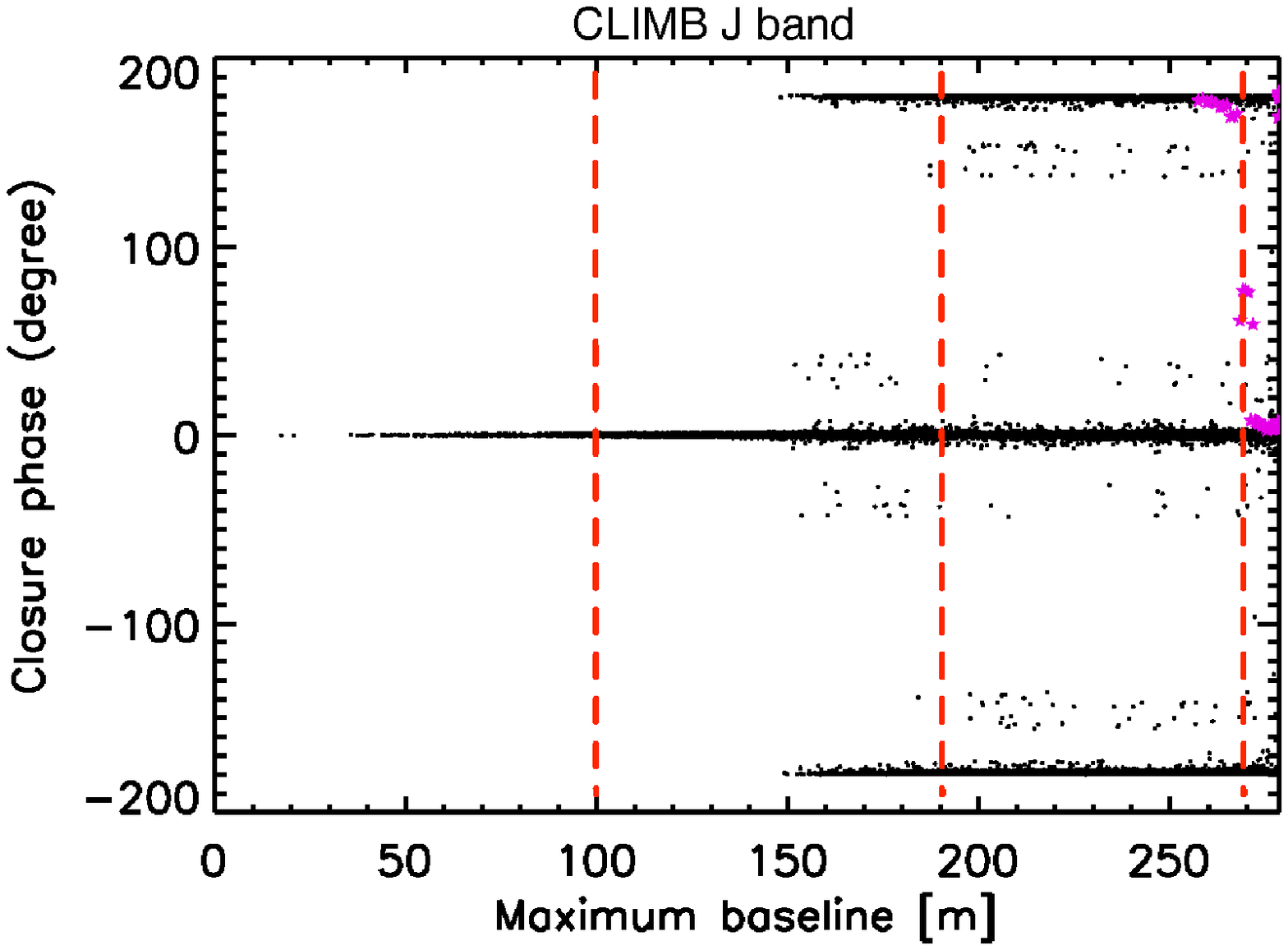}
              \includegraphics[width=0.32\hsize]{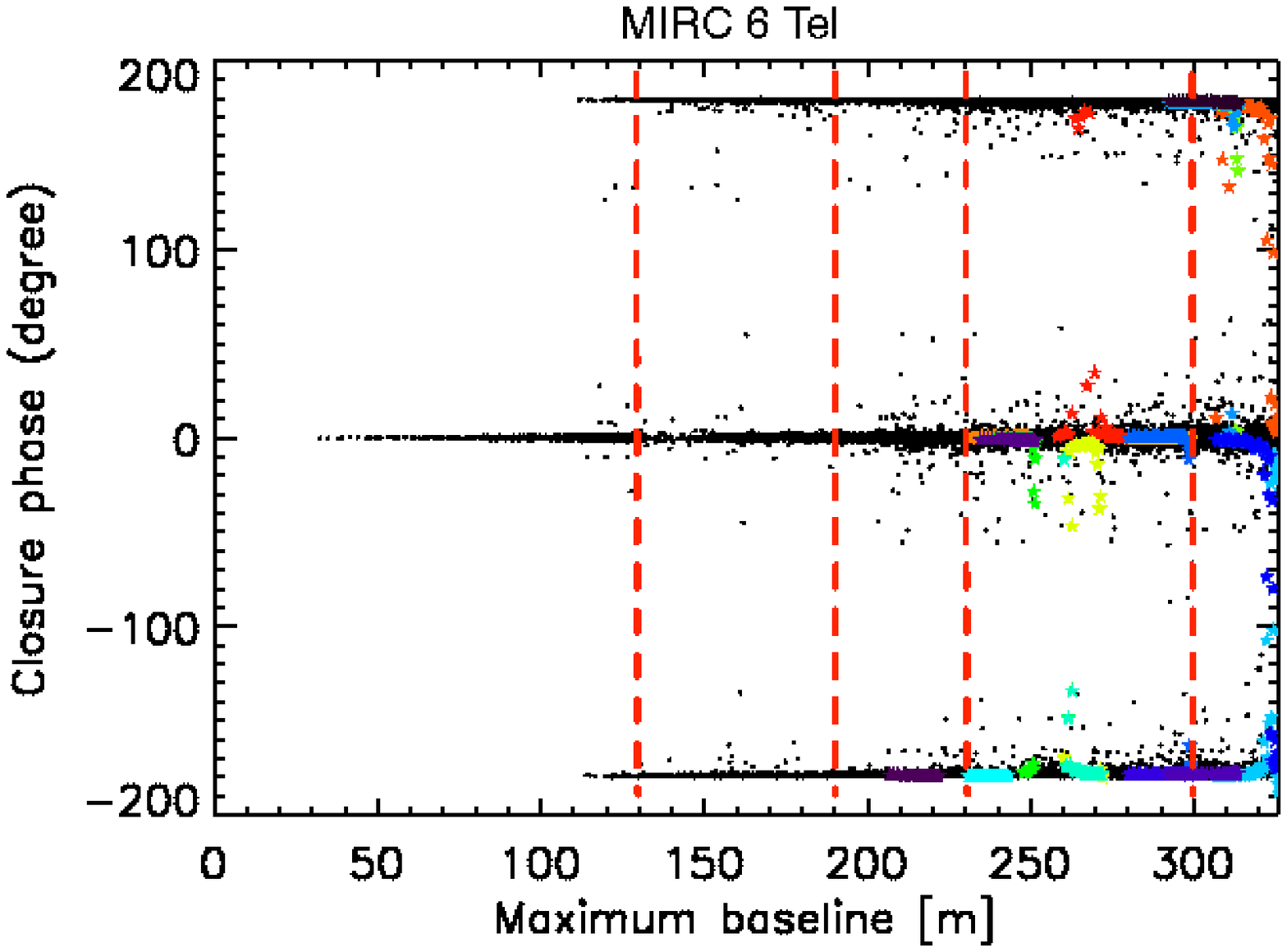}\\
 	   \includegraphics[width=0.32\hsize]{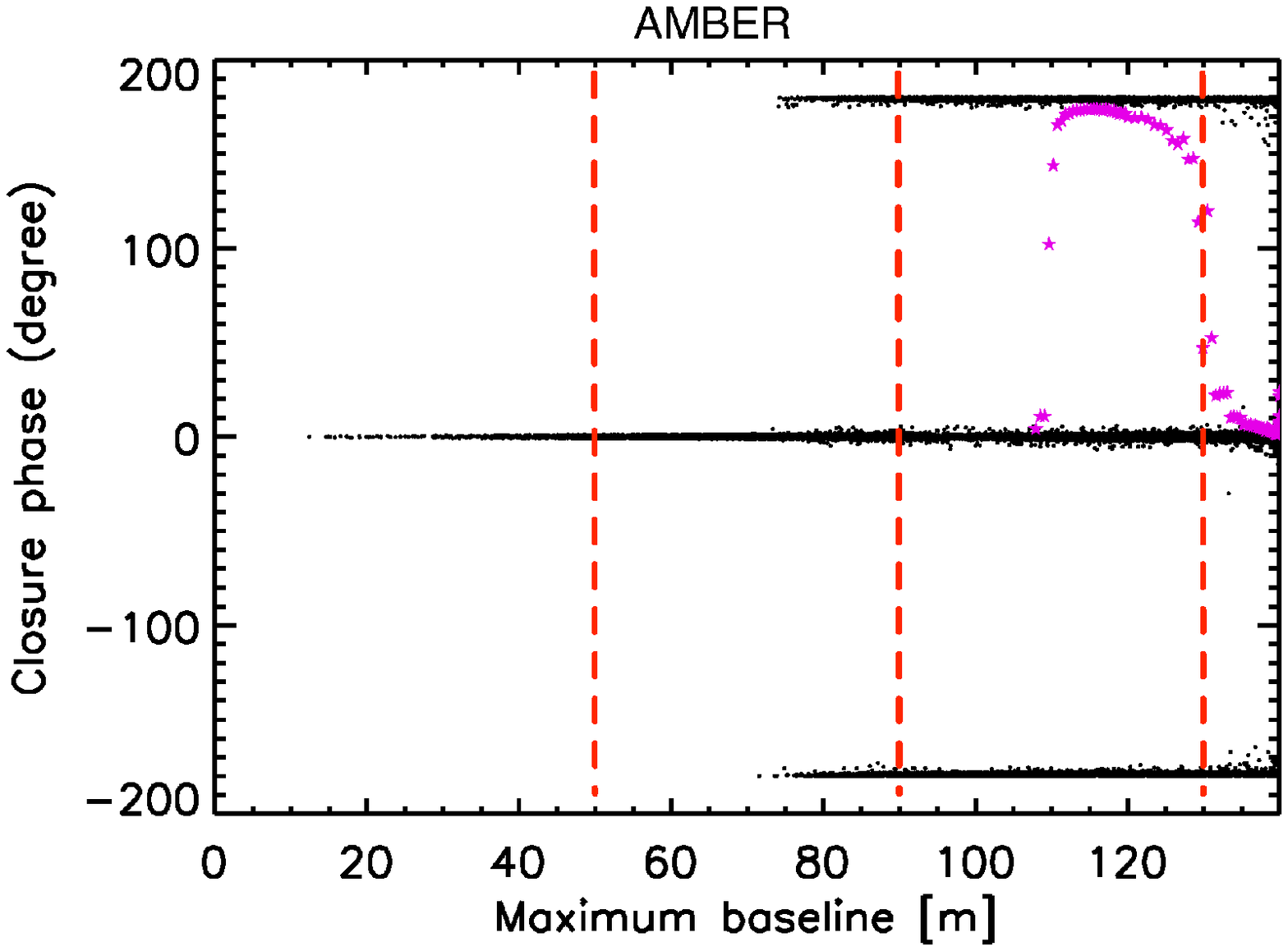}
              \includegraphics[width=0.32\hsize]{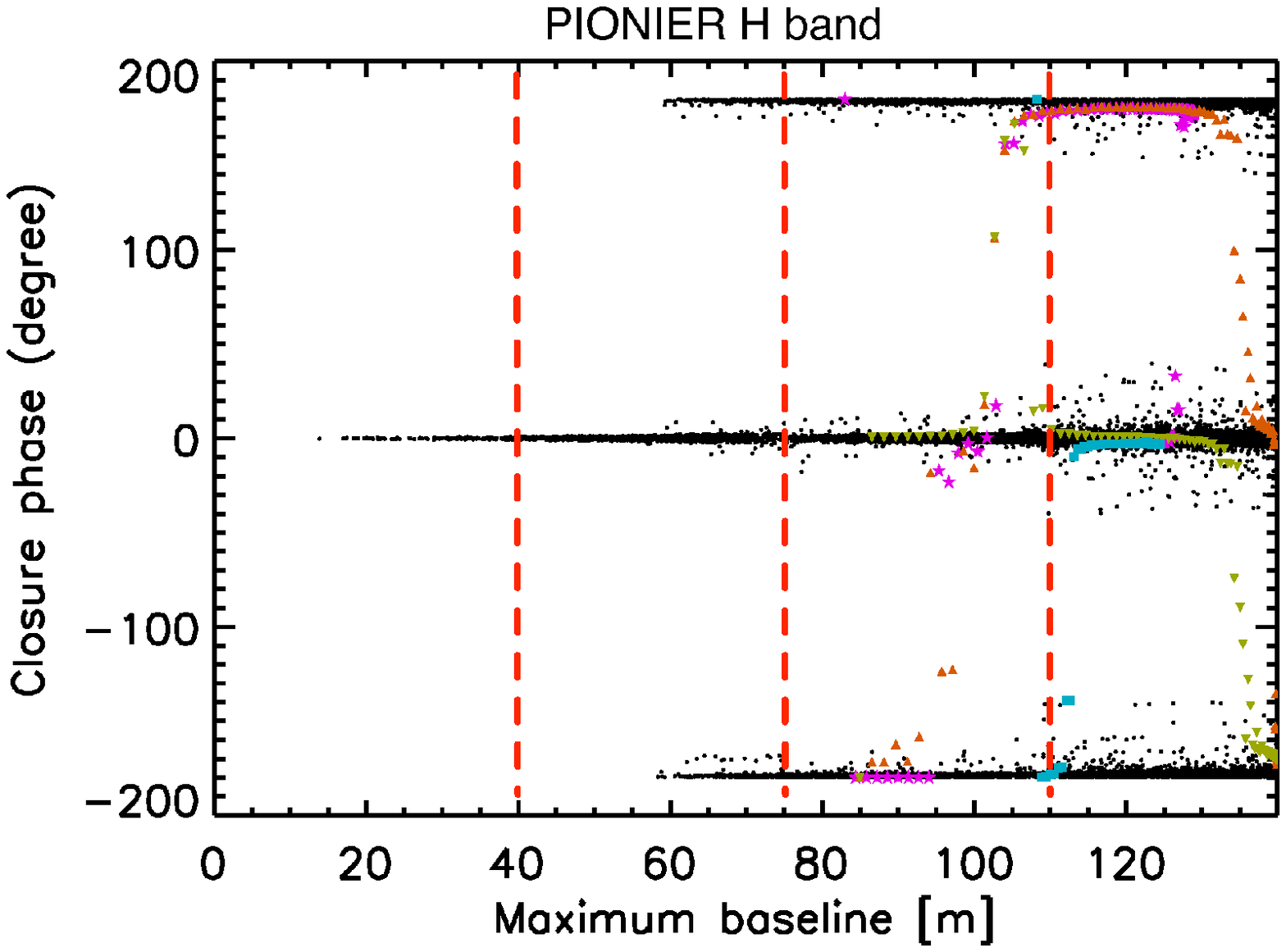}
   	\includegraphics[width=0.32\hsize]{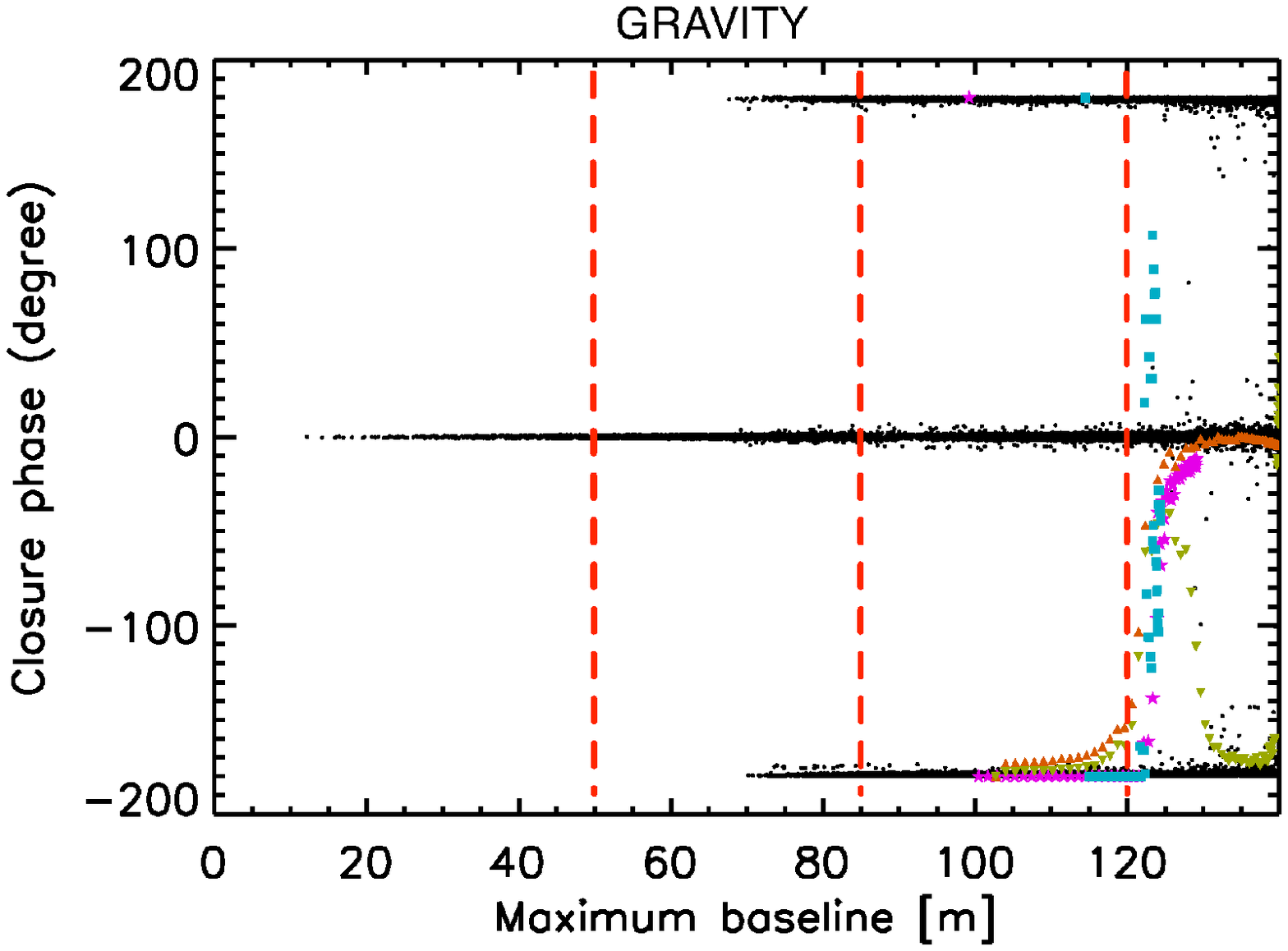}\\
              \includegraphics[width=0.32\hsize]{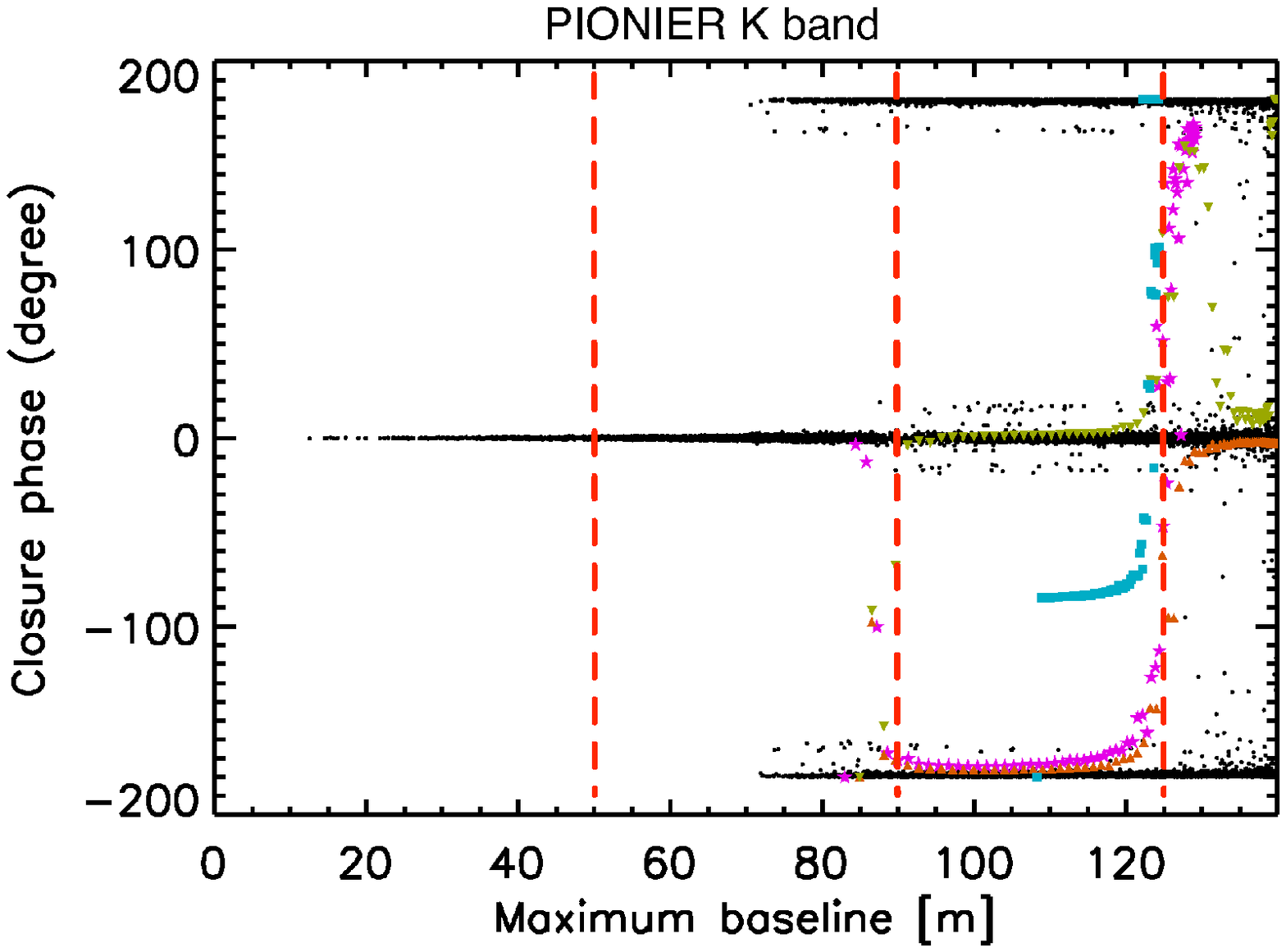}
      \includegraphics[width=0.32\hsize]{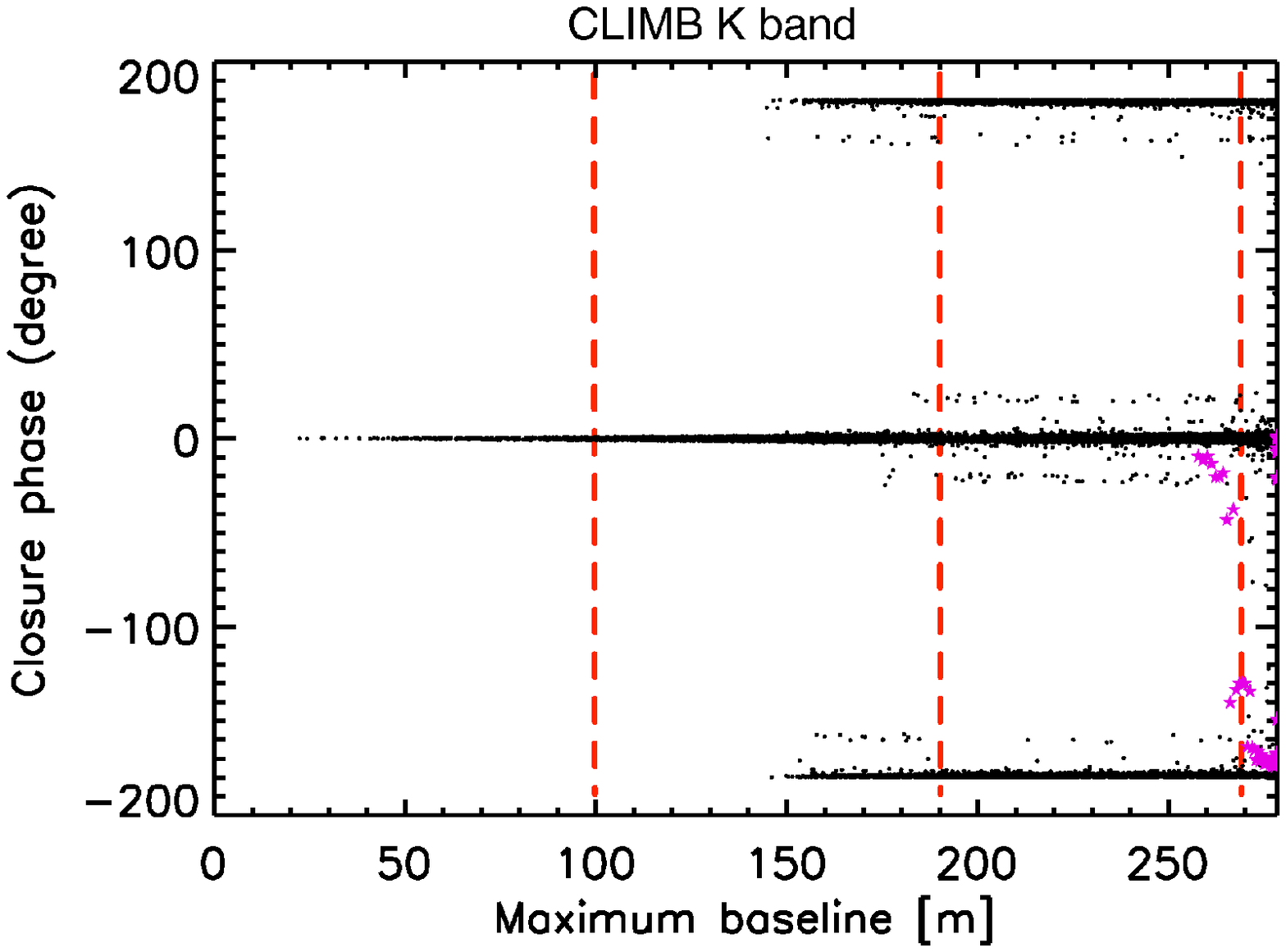}
              \includegraphics[width=0.32\hsize]{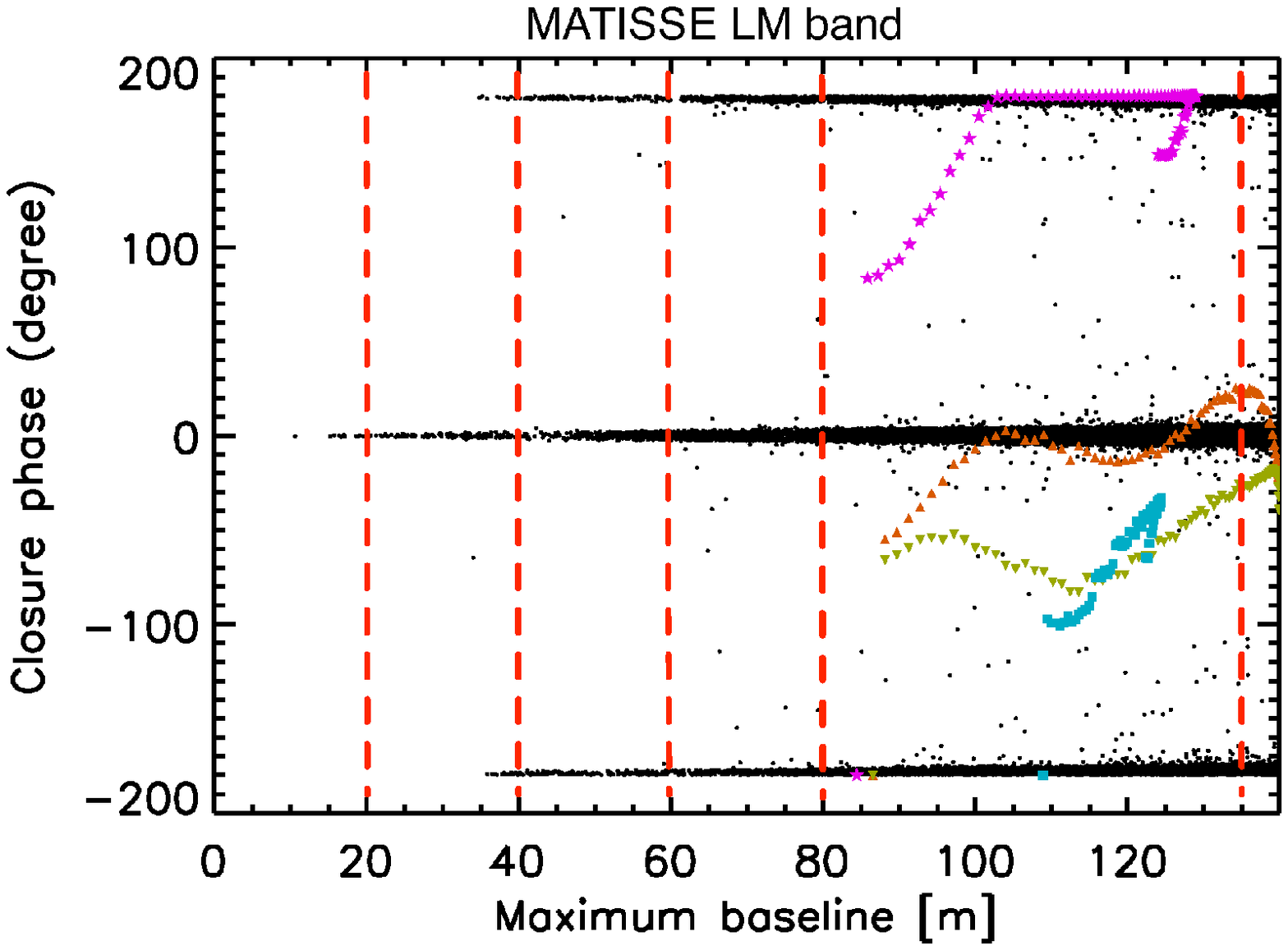}\\
        \end{tabular}
      \caption{Scatter plot of closure phases of 20000 random baseline triangles (black dots) as a function of the maximum linear extension corresponding to the configuration chosen for each instrument of Table~\ref{instruments} and for the RHD simulation of the Sun (Table~\ref{simus}). The colored symbols over-plotted display the closure phases for the configuration chosen (see the UV-planes of Fig.~\ref{uvplan}) and the vertical dashed red lines give the approximative positions  of the different lobes.}
        \label{closure}
   \end{figure*}

The stellar surface asymmetries in the brightness distribution, due to either convection-related structures or a faint companion, affect closure phases. The sum of three phases around a closed triangle of baselines is the closure phase: this procedure removes the atmospheric contribution, leaving the phase information of the object visibility unaltered \citep{2007NewAR..51..604M,2003RPPh...66..789M}. Closure phases have the main advantage of being uncorrupted by telescope-specific phase errors, including pointing errors, atmospheric piston, and longitudinal dispersion due to air and water vapor \citep{2012A&A...541A..89L}. Closure phase errors, when known, are reported in Table~\ref{instruments}.\\
However, owing to the sparse structure of the point spread function associated with the diluted aperture of an interferometer (see Fig.~\ref{uvplan}), the position and the morphology of these surface inhomogeneities depend on their relative orientation and on the interferometric baselines. 
  
Figure~\ref{images} shows irregular stellar surfaces with convection-related surface structures, whose sizes depend on the stellar parameters of the simulations. There are pronounced centre-to-limb variations in the optical (VEGA to NPOI instruments) while these are less noticeable in the infrared. This is mainly due to the differences in Planck functions in the optical range and in the infrared region. \\
Starting from these synthetic images, we followed the method described in \cite{2009A&A...506.1351C} to calculate the discrete complex Fourier transform $\mathcal{F}$ for each image, with particular interest only in the closure phases. From the OIFITS files for each instrument we know the set of all baseline vectors (i.e., UV-plane) of all the telescopes (see Sect.~\ref{interfere}) and we matched their frequencies in the UV-plane with the corresponding points in the Fourier transform of the synthetic images. The phase for each telescope is $\tan\varphi = \Im(\mathcal{F})/\Re(\mathcal{F})$, where $\Im(\mathcal{F})$ and $\Re(\mathcal{F})$ are the imaginary and real parts of the complex number $\mathcal{F}$, respectively. 
Finally, the closure phase is the sum of all phase differences between closed triangles of telescope baselines: e.g. for 3 telescopes: $CP(1-2-3)=\Phi_{1-2}+\Phi_{2-3}+\Phi_{3-1}$, where $CP(1-2-3)$ is the closure phase, $\Phi_{1-2}$ is the $\arctan$ of the Fourier phases $\tan\varphi$ for the telescopes 1-2.\\
In our survey, we used the setup of different instruments (Table~\ref{instruments}) characterised by a number of telescopes ($N$) varying from 3 to 6. \cite{2003RPPh...66..789M} showed that the possible closed triangles of baselines (i.e, one closed triangle gives one closure phases) is ${N\choose 3}=\frac{\left(N\right)\left(N-1\right)\left(N-2\right)}{\left(3\right)\left(2\right)}$, but the independent Fourier phases are given by ${N\choose 2}=\frac{\left(N\right)\left(N-1\right)}{\left(2\right)}$, and thus, not all the closure phases are independent but only ${N-1\choose 2}=\frac{\left(N-1\right)\left(N-2\right)}{\left(2\right)}$. \\
The number of independent closure phases is always less than the number of phases one would like to determine, but the percent of phase information retained by the closure phases improves as the number of telescopes in the array increases \citep{2003RPPh...66..789M}.

To sum up:
\begin{itemize}
\item with 3 telescopes one obtains 1 closed triangle of baselines (i.e., 1 closure phase), 3 Fourier phases, and 1 independent closure phase
\item with 4 telescopes one obtains 4 closed triangles of baselines  (4 closure phases), 6 Fourier phases, and 3 independent closure phases
\item with 6 telescopes one obtains 20 closed triangles of baselines (20 closure phases), 15 Fourier phases, and 10 independent closure phases
\end{itemize}

\cite{2012A&A...540A...5C, 2010A&A...524A..93C} demonstrated that, in the case of Procyon and K-giant stars, the synthetic visibility curves produced by the RHD simulations are systematically different from spherical symmetric modeling, with an impact on the radius, effective temperature, and departures from symmetry. This was noticeable at higher spatial frequencies and mostly affecting the signal of the closure phases. The authors interpreted this as the signature linked to the convection-related surface structures. Starting from these remarks, we decided to concentrate our survey study only on the closure phases.

 \begin{figure*}
   \centering
   \begin{tabular}{c}
              \includegraphics[width=0.98\hsize]{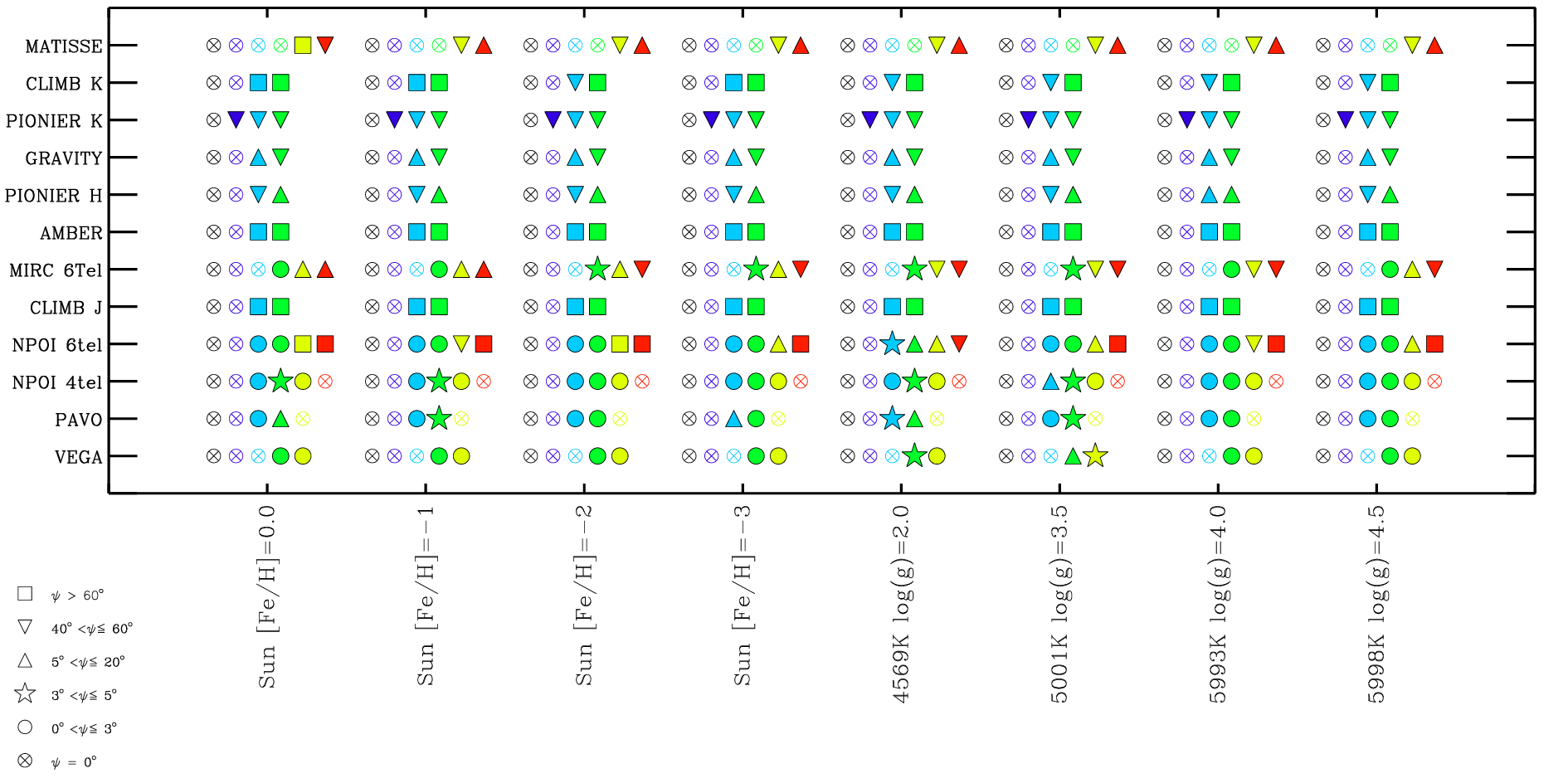}
        \end{tabular}
      \caption{Differences, $\psi$, between the closure phases and zero or $\pm\pi$ (i.e., departure from axisymmetric case) for all the simulations of Table~\ref{simus} (horizontal axis) and all the instruments of Table~\ref{instruments} (vertical axis). $\psi$ has been averaged over the spatial frequencies corresponding to the different lobes spanned by the instruments configuration (see text) and, for each simulation, 6 lobes are displayed: black for the 1st lobe, violet for the 2nd, light blue for the 3rd, green for the 4th, yellow for the 5th, and red for the 6th. Only the lobes spanned in the UV-planes of Fig.~\ref{uvplan} are plotted. The symbols correspond to different values, in degrees. }
              \label{closurebis}
   \end{figure*}

Figure~\ref{closure} displays closure phases deviating from the axisymmetric case that are particularly occurring in optical filters, where the dispersion is larger (e.g., VEGA, NPOI, and PAVO instruments). Depending on the instruments and spatial frequency spanned, the departures from symmetry may be large or not. However, it is apparent that the convection-related surface structures have a signature on the closure phases. \\

The characterisation of the granulation signature is analyzed in Fig.~\ref{closurebis}, where the departure from axisymmetric case (i.e., closure phase different from zero or $\pm\pi$) of all the  RHD simulations and interferometric instruments is displayed. We proceed as follows:

\begin{itemize}
\item for all the instruments, we determined the difference, $\psi_i$, between the  synthetic closure phases and the axisymmetric values zero or $\pm\pi$ from all spatial frequency $i$;
\item based on the UV coverage of Fig.~\ref{uvplan} and the synthetic visibilities, we identified the frequency limits of the lobes (vertical dashed red lines of Fig.~\ref{closure});
\item we averaged $\bar{\psi_i}$ over the frequencies $i$ falling inside the lobe's limits;
\item in case of multiple closed triangles (i.e., multiple values of closure phases such as for instruments working with 4 or 6 telescopes), we selected the largest $\bar{\psi_i}$. 
\end{itemize}

It is remarkable that all the simulations show departure from the axisymmetric case at all the wavelengths. At least for the chosen configurations, it is difficult to determine clear differences among the stellar parameters and, in particular, for the different metallicities of the solar simulations. In addition to this, it must be noted that the averaged $\bar{\psi_i}$ may smooth out 
the differences even if these observables are useful for pointing out the signature of the convection-related surface structures. \\
The levels of asymmetry and inhomogeneity of stellar disk images reach very high values of several tens of degrees with stronger effects from 3rd visibility lobe on. 
In this work we assumed precise values for the arbitrary observations (see Fig.~\ref{uvplan}). To estimate the baseline needed for other stellar sizes, the following equation can be used to retrieve second zero of the visibility curve (i.e., the third lobe):
\begin{equation}
B[\rm{m}]  = 2.23 \cdot \frac{\lambda}{\theta[rad]} = 2.23 \cdot \frac{\lambda\cdot 206265 \cdot 1000}{\theta[mas]}
\end{equation}
where $\theta$ is the apparent diameter, $\lambda$ is the wavelength in meters, and $B$ is the baseline in meters. For example, a sun observed at wavelength 0.7 $\mu{\rm m}$ (e.g., VEGA instrument) with $\theta$= [0.5, 2, 5, 10] mas the third lobe will be probed with a baselines of about [645, 165, 65, 33] m. The same example in the H band 1.6 $\mu{\rm m}$ (e.g., PIONIER instrument) gives baselines of about [1472, 368, 148, 74] m, respectively; and for the L band at 30 $\mu{\rm m}$ with baselines of [27600, 2900, 2760, 1380] m, respectively.\\

Finally, to estimate the distance of the observed star the following equation can be used: 
\begin{equation}
d~[{\rm pc}] = \frac{R[{\rm R}_\odot] \cdot  9.305}{\theta[mas]} 
\end{equation}
where $d$ is the distance of the observed star in parsec, $R[{\rm R}_\odot]$ is the radius of the star in solar radii, and $\theta$ is the apparent diameter.
The optical and the near infrared wavelengths are more affordable in term of baseline lengths because they are less than 400 meters for stellar sizes larger than 2 mas, it is however more complicated for the mid-infrared wavelengths where the baseline lengths become kilometric. The signature on the closure phases can be evaluated by accumulating observations at short and long baselines. This can be ensured by the fact that the instrumental errors on closure phases is much smaller than the expected closure phase departures (see Table~\ref{instruments}). It is however important to note that probing high frequencies, the signal to noise ratio of the measurements would be very low due to low fringe visibilities, greatly deteriorating the closure phase precision and affecting the instrument capability.

\section{Applications} \label{realsection}
          
\subsection{Study of the granulation on two real targets}             
          
          In previous section we showed that the detection of closure phase departures from symmetry needs stars resolved up to the 3rd and 4th lobes, as well as in some cases, the 5th or 6th lobes. This reduces the number of targets that can be observed with the actual interferometric baselines but foresees the need for an extension of the next generation interferometric infrastructures.\\
          
\begin{table}
 \small
\centering
%\begin{minipage}[t]{\textwidth}
\caption{Reference targets and associated RHD simulations of Table~\ref{simus}.}             % title of Table
\label{realstars}      % is used to refer this table in the text
\centering                          % used for centreing table
\renewcommand{\footnoterule}{} 
\begin{tabular}{c c c c c }        % centreed columns (4 columns)
\hline\hline                 % inserts double horizontal lines
Name  & Spectral  & angular  &  RHD  \\
	    & type          & diameter [mas] &  [$<T_{\rm{eff}}>$] \\        
 	    \hline
Beta Com (HD~114710) & G0V &  1.1\tablefoottext{a} & Sun   \\
Procyon (HD~62421)    & F5IV 5.4 & 5.4\tablefoottext{b} & 5993.42 \\ 

\hline\hline                          % inserts single horizontal line
\end{tabular}
\tablefoottext{a}{\cite{2005A&A...431..773R}}	
\tablefoottext{b}{\cite{2012A&A...540A...5C}}
%\end{minipage}
\end{table}

           In this section, we performed the closure phase analysis for two real targets: Beta Com and Procyon (Table~\ref{realstars}). Beta Com has been chosen for its smaller angular diameter so as to illustrate observations in the visible while the large diameter of Procyon ensures a good UV coverage in the infrared. As before, we prepared adapted OIFITS files for each instrument using real observability coverage.
               
    \begin{figure}
   \centering
   \begin{tabular}{c}
        \includegraphics[width=1.0\hsize]{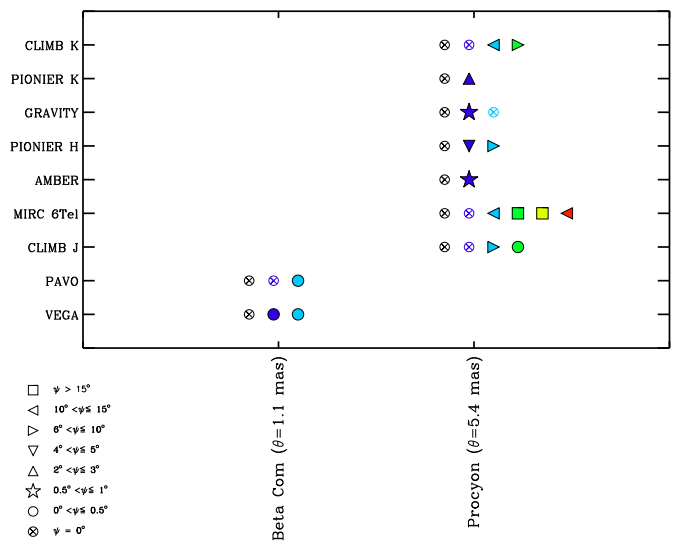}\\
          \end{tabular}
      \caption{Same as in Fig.~\ref{closurebis} but for two real targets: Beta Com and Procyon (see Table~\ref{realstars} for the associated RHD simulation). MATISSE and NPOI do not probe frequency larger than the first lobe and have not been reported. The size of the circles correspond to different value, in degrees. Note that the values of $\psi$ are smaller than in Fig.~\ref{closurebis}. The crossed-circle means that there are no detected departures}
        \label{realstarsfig}
   \end{figure}
        
            Fig.~\ref{realstarsfig} displays smaller departures in closure phases with respect to the configurations taken in Fig.~\ref{closurebis}: in general all the instruments (except for MATISSE and NPOI, which do not probe frequencies larger than the first lobe) show closure phases departures ($\psi$) of the order of few degrees with largest values of the order of $\sim16^\circ$ (to be compared with instrumental errors of Table~\ref{instruments}). \\
          As we only considered, in this instance, one single target for all the visible instruments, and another for the infrared instruments, we can directly compare the results among the different instruments. Both PAVO and VEGA return values lower than 0.5$^\circ$ with a closure phase signal starting from the 2nd lobe for VEGA. More interesting is the infrared region with VLTI's instruments showing departures already from the 2nd lobe: AMBER and GRAVITY with values lower than 0.8$^\circ$, PIONIER in the H band with values of 4.3$^\circ$ (2nd lobe) 6.4$^\circ$ (3rd lobe), and PIONIER in the K band with 2.9$^\circ$ (2nd lobe). CHARA's instruments do not show departures on the 2nd lobe but they probe higher frequencies up to the 6th lobe with MIRC (13.8$^\circ$, 15.3$^\circ$, 16.4$^\circ$, 13.1$^\circ$ for the 3rd, 4th, 5th, 6th lobe), CLIMB J band (6.5$^\circ$, 0.2$^\circ$, 5.1$^\circ$ for the 3rd, 4th, 5th lobe), and CLIMB K band (12.3$^\circ$, 6.3$^\circ$ for the 3rd, 4th lobe). \\
           The actual instruments and telescopes, with the errors on closure phases reported in Table~\ref{instruments}, allow, in principle, the detection of the granulation. The closure phase signal is already more pronounced in the infrared for the 2nd lobe and may be detected with very good weather and instrumental conditions but it is certainly easier to detect from the 3rd lobe on. The long baseline set of CHARA telescopes is more advantageous as higher frequencies are probed. Moreover, MIRC instrument with 6 telescope recombination is the most appropriate instrument as it combines good UV coverage and long baselines. Our analysis is based on the assumption of very good conditions, but, in the context of non-zero exposure times and the presence of atmospheric turbulence, the accuracy on closure phase measurements is also affected by piston noise. The statistical uncertainty on the closure phases depends on the atmospheric conditions with the piston noise contribution decreasing with the square root of the integration time \citep{2012A&A...541A..89L}. Optimal observing strategy could however be defined to reach the needed accuracy as already demonstrated by some of instruments of Table~\ref{instruments}. 
          
          \subsection{Transiting planet}   

\begin{figure*}
   \centering
%  \begin{tabular}{c@{\hspace{0.9truecm}}c@{\hspace{0.9truecm}}c}
 \begin{tabular}{rcl}
	     \includegraphics[width=0.33\hsize]{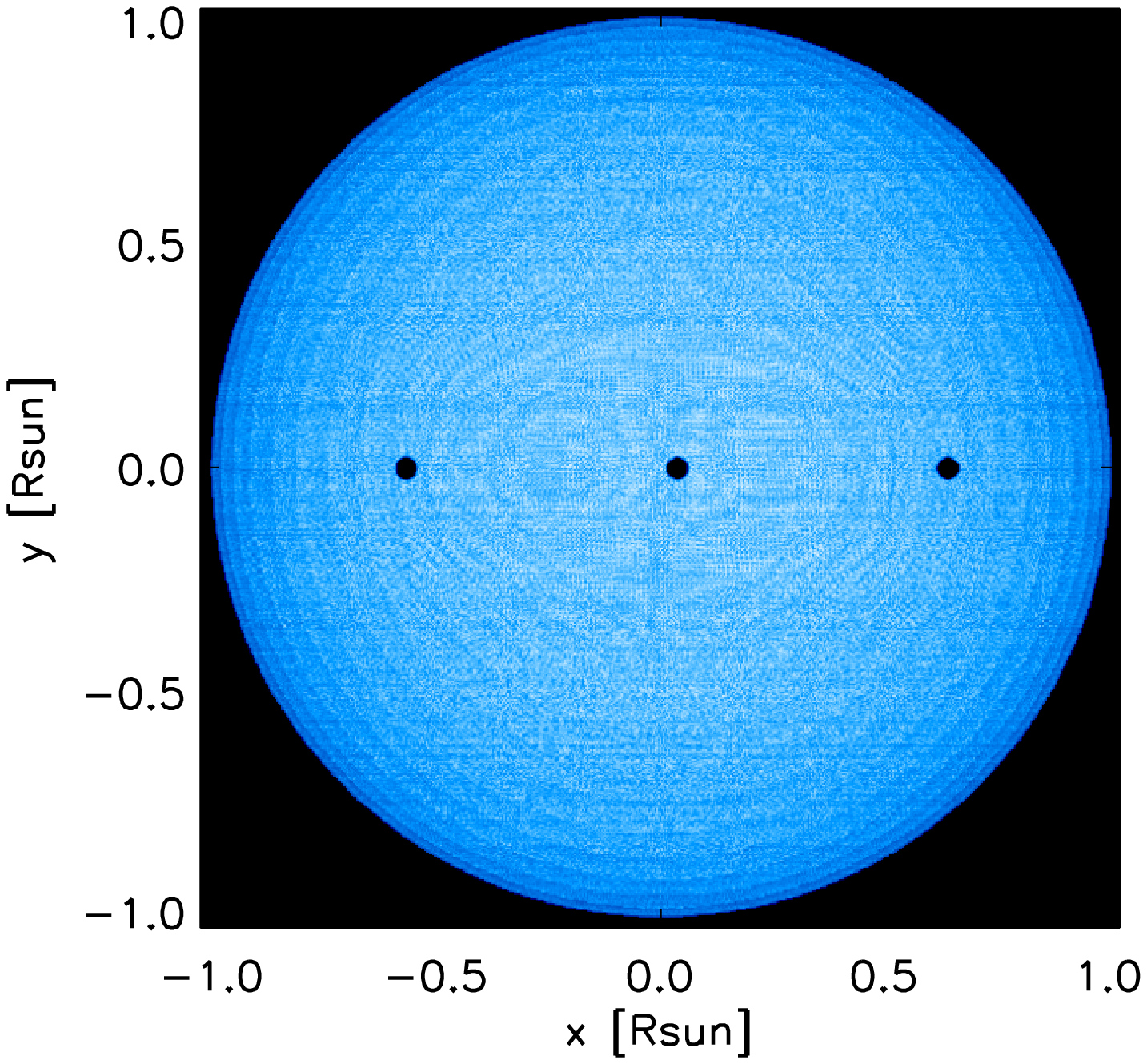}
              \includegraphics[width=0.33\hsize]{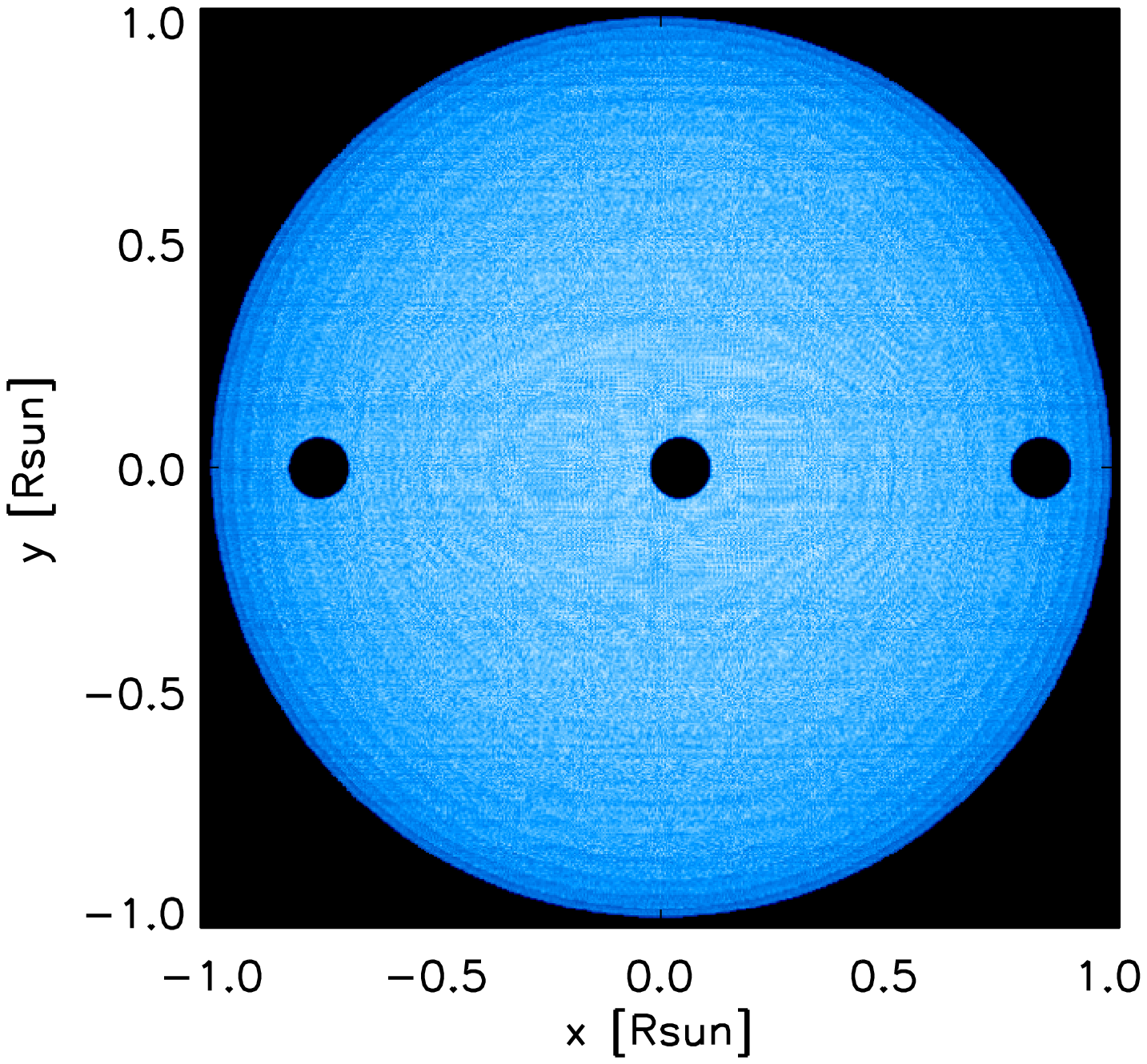}
              \includegraphics[width=0.33\hsize]{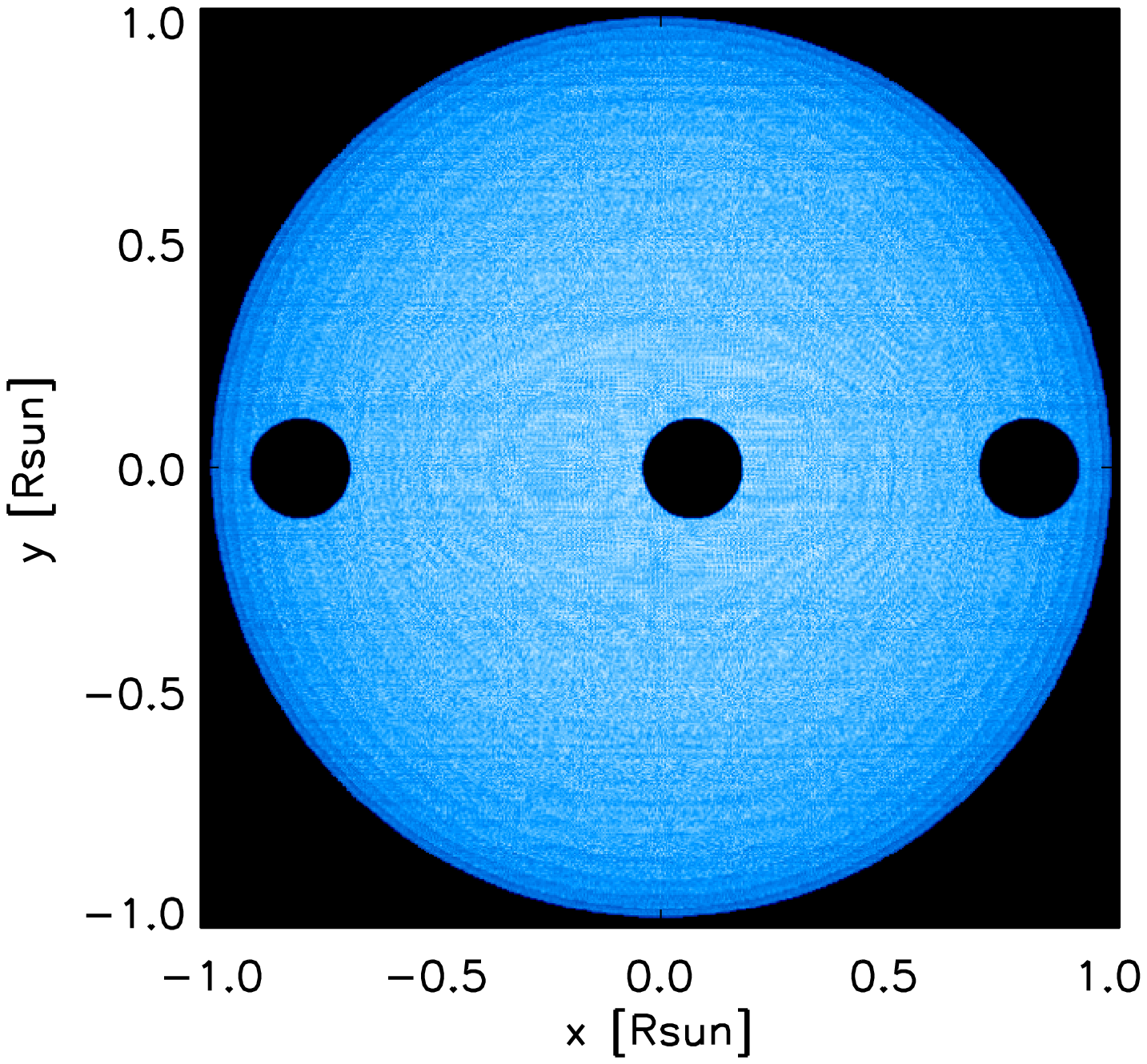}\\
         \end{tabular}
      \caption{Synthetic stellar disk images in PIONIER H band together with three planet transiting phases (black color) for the Sun of Table~\ref{simus}. The prototypes of planet are from Table~\ref{planets} and are: Kepler-11 f prototype planet (left column), HD 149026 b prototype (central column), and CoRoT-14 b prototype (right column). The purpose is not to reproduce the exact conditions of the planet-star system already detected but to have a statistical approach on the interferometric signature for different stellar parameters hosting planets with different sizes.}
        \label{transit}
   \end{figure*}

\begin{figure*}
   \centering
   \begin{tabular}{cccc}
\includegraphics[width=0.24\hsize]{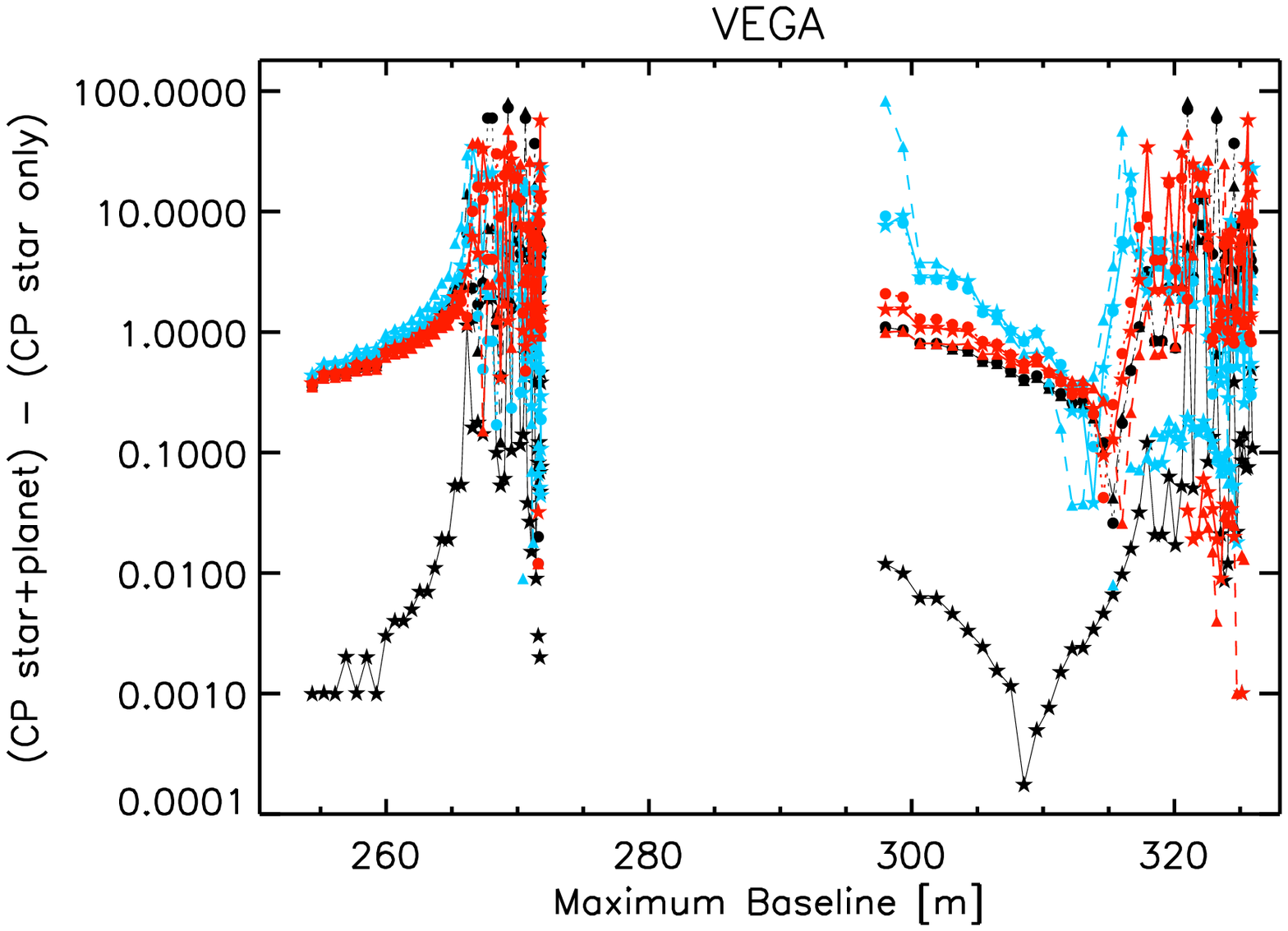}	
\includegraphics[width=0.24\hsize]{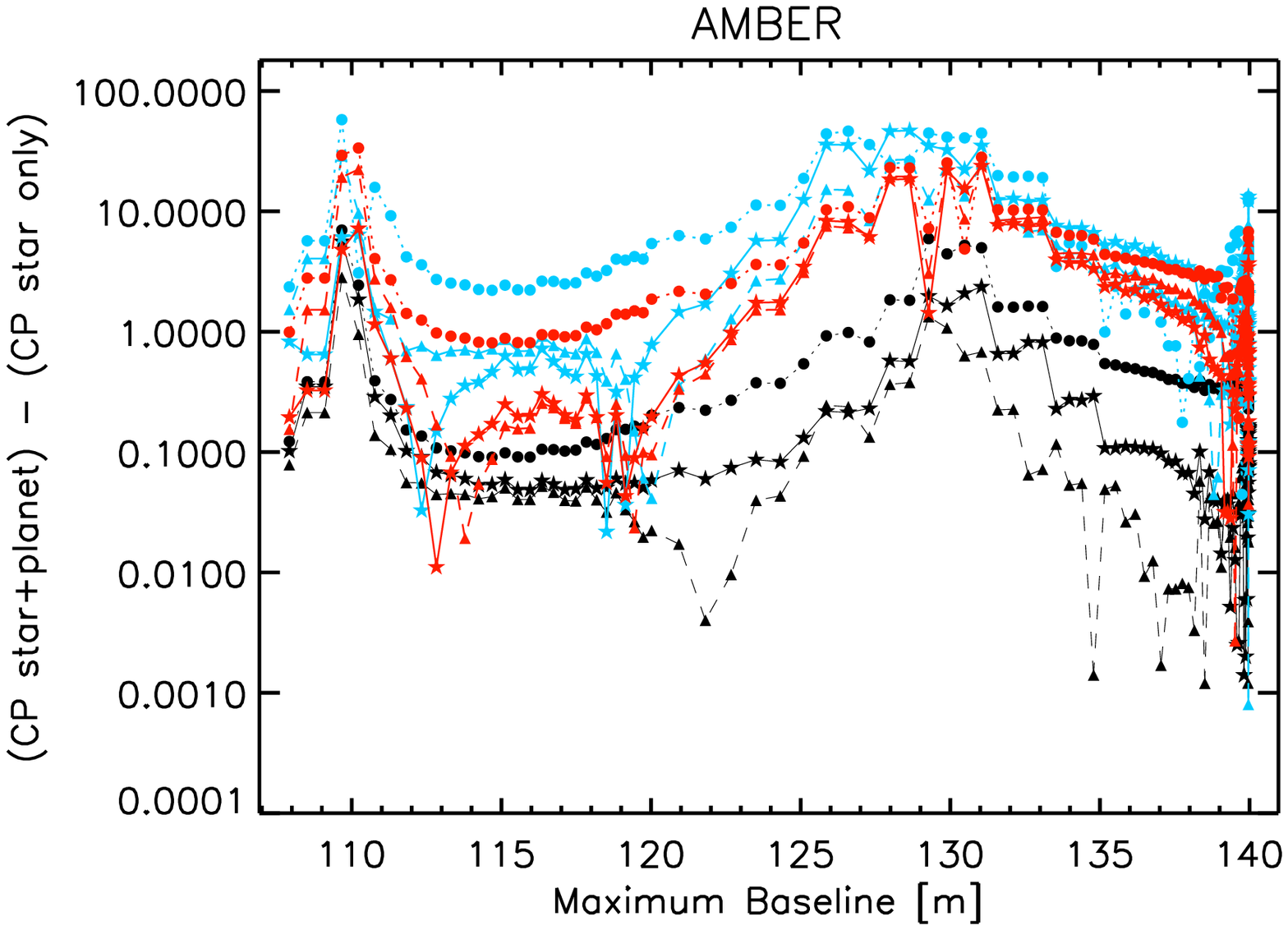}
\includegraphics[width=0.24\hsize]{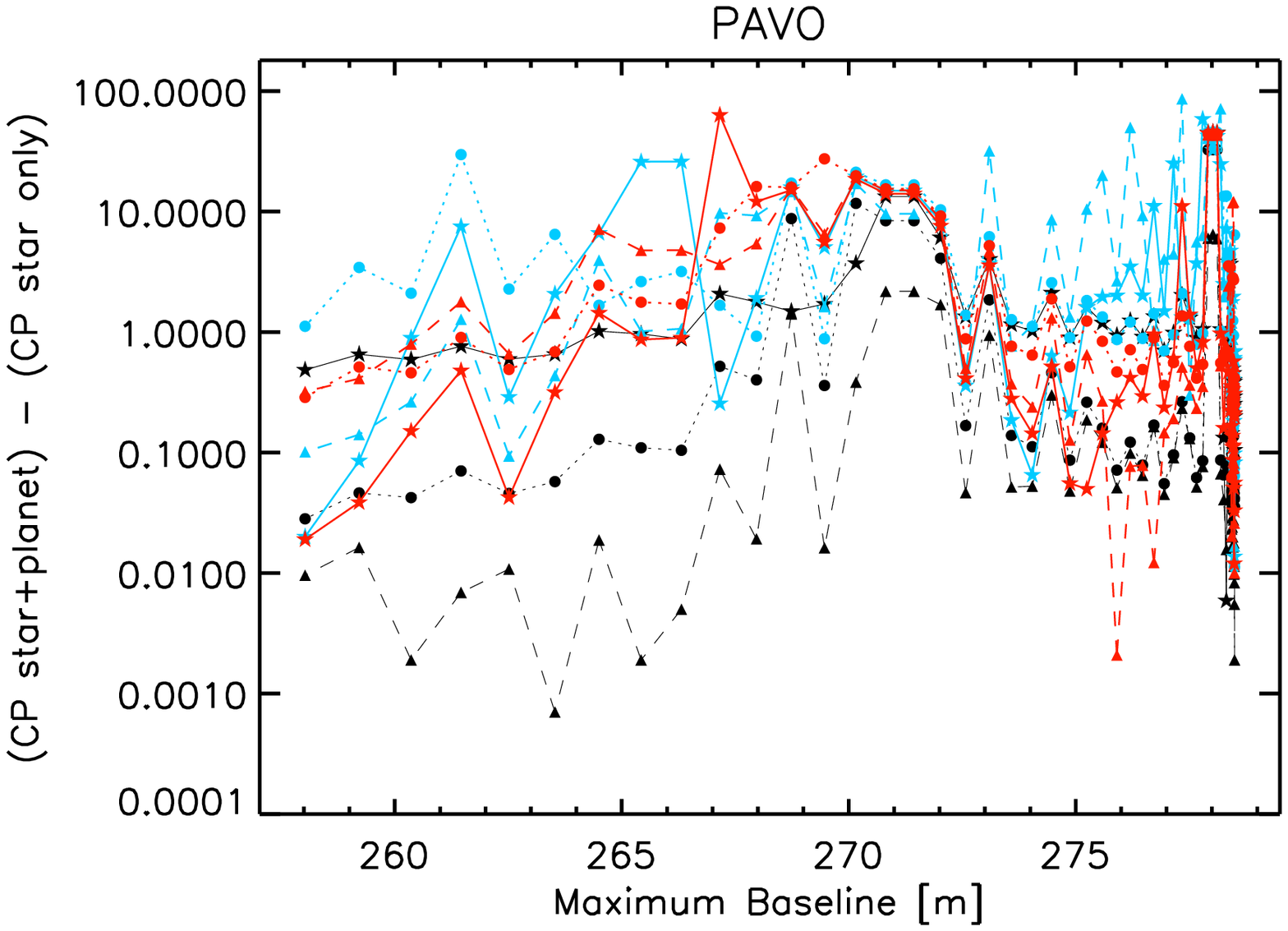}	
\includegraphics[width=0.24\hsize]{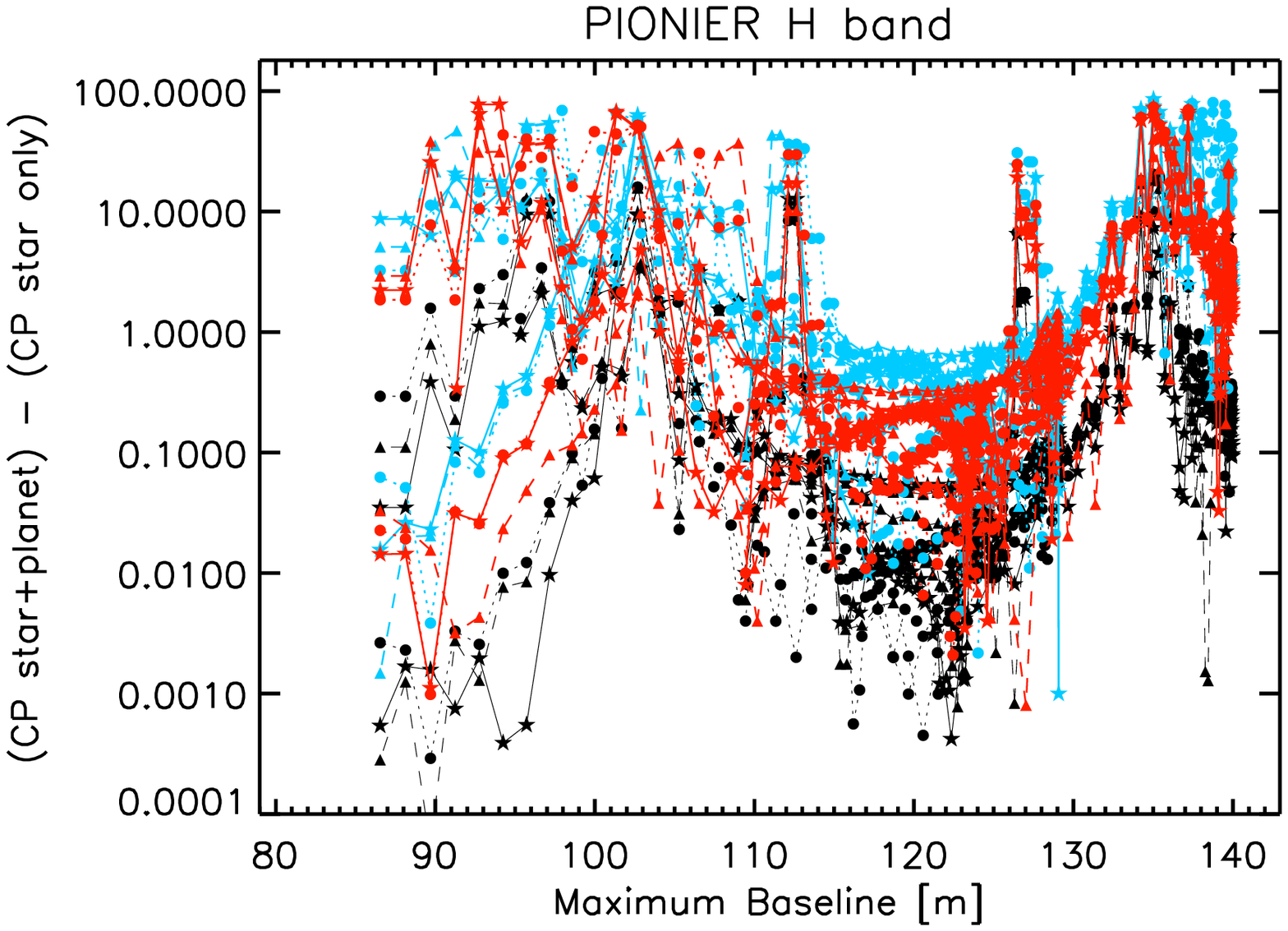}\\
\includegraphics[width=0.24\hsize]{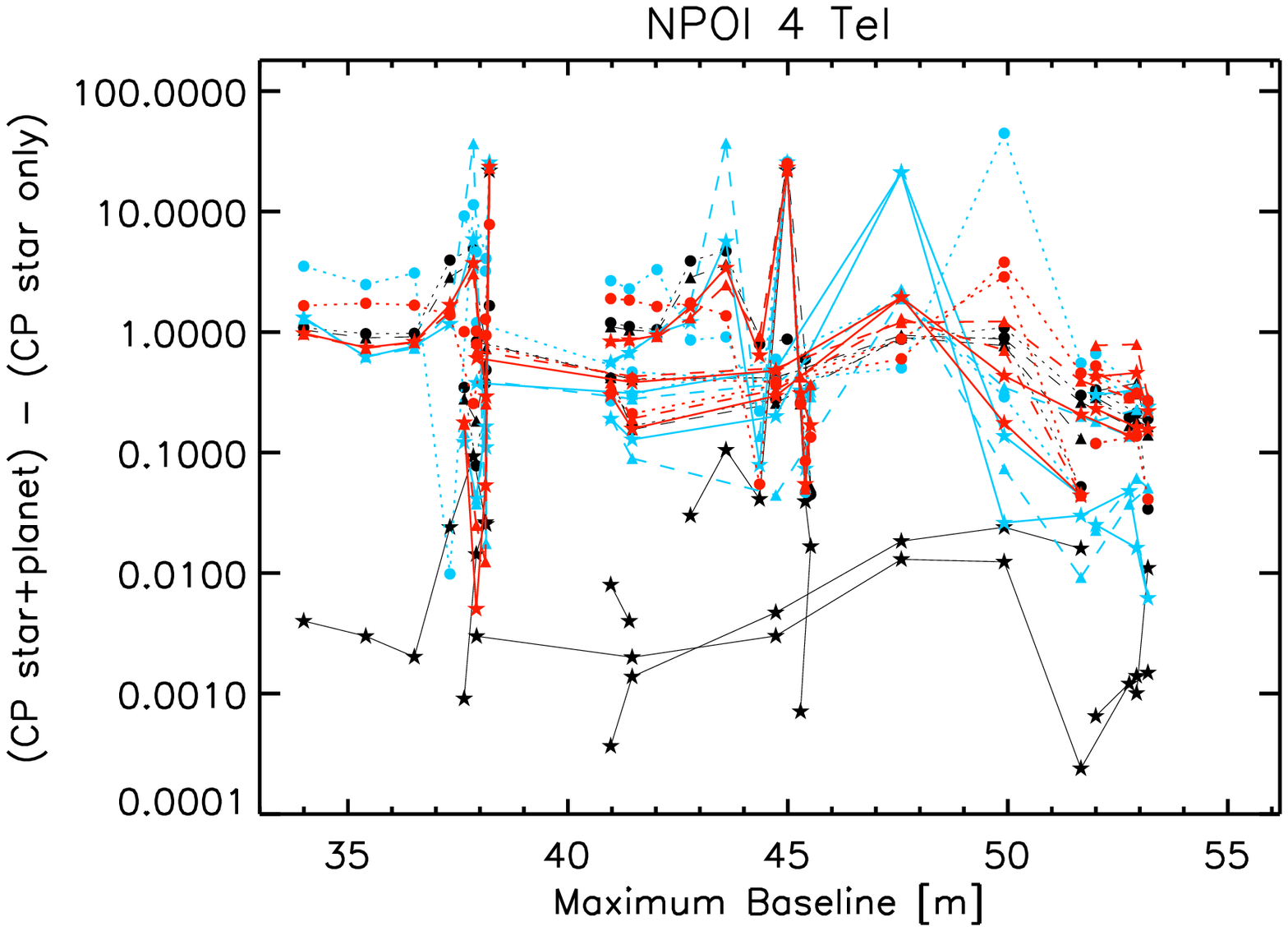}	
\includegraphics[width=0.24\hsize]{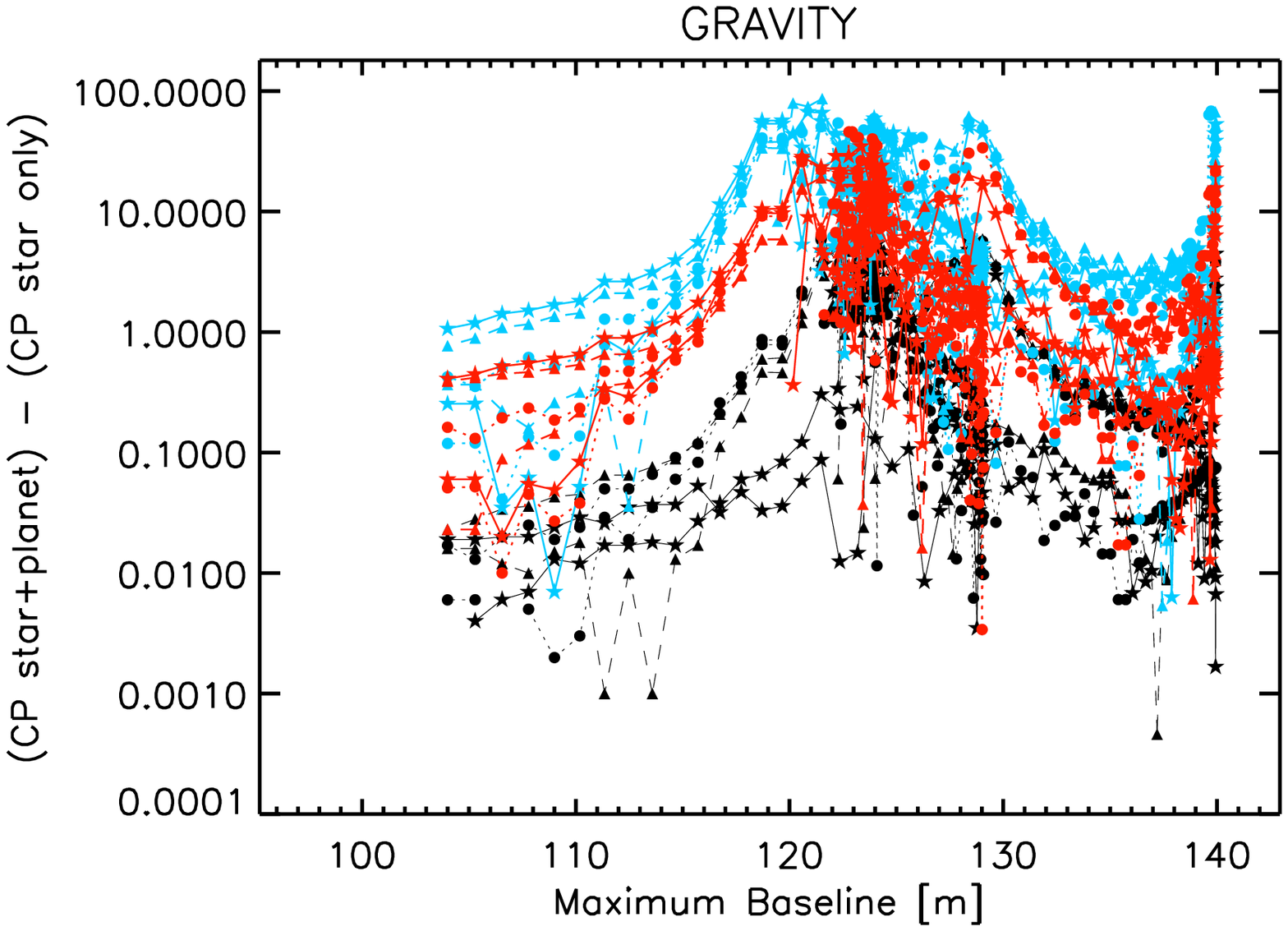}
\includegraphics[width=0.24\hsize]{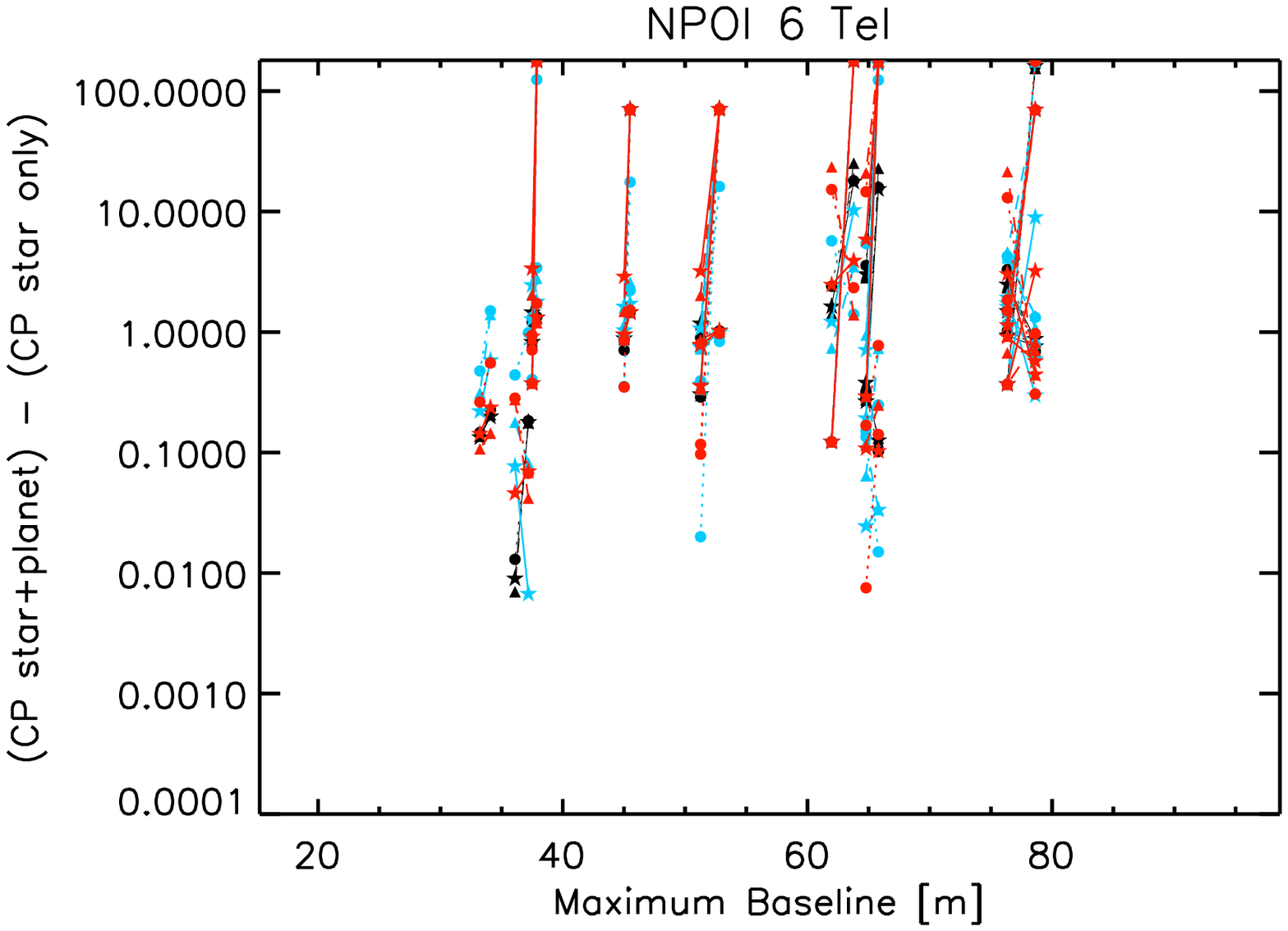}	
\includegraphics[width=0.24\hsize]{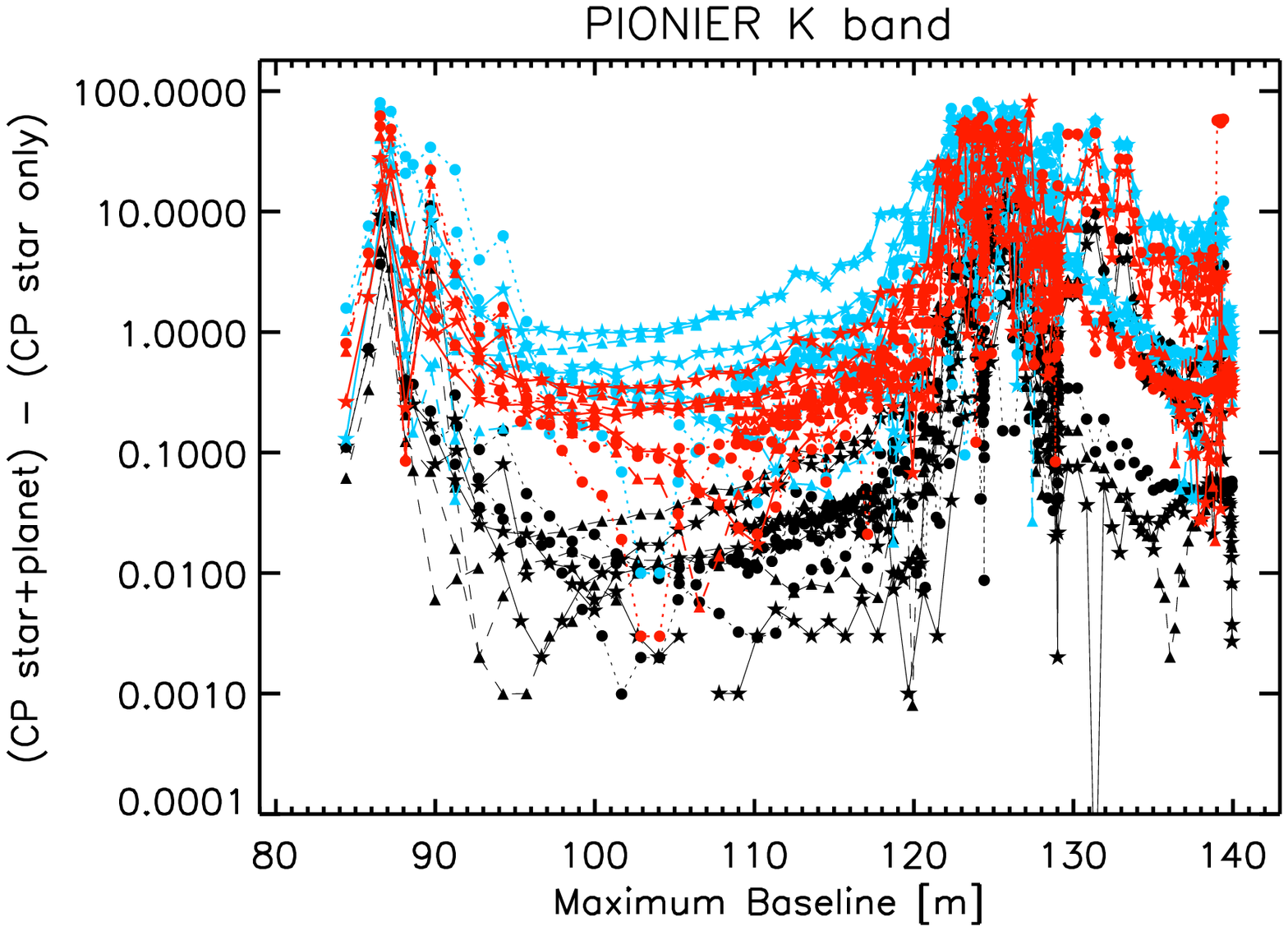}\\
\includegraphics[width=0.24\hsize]{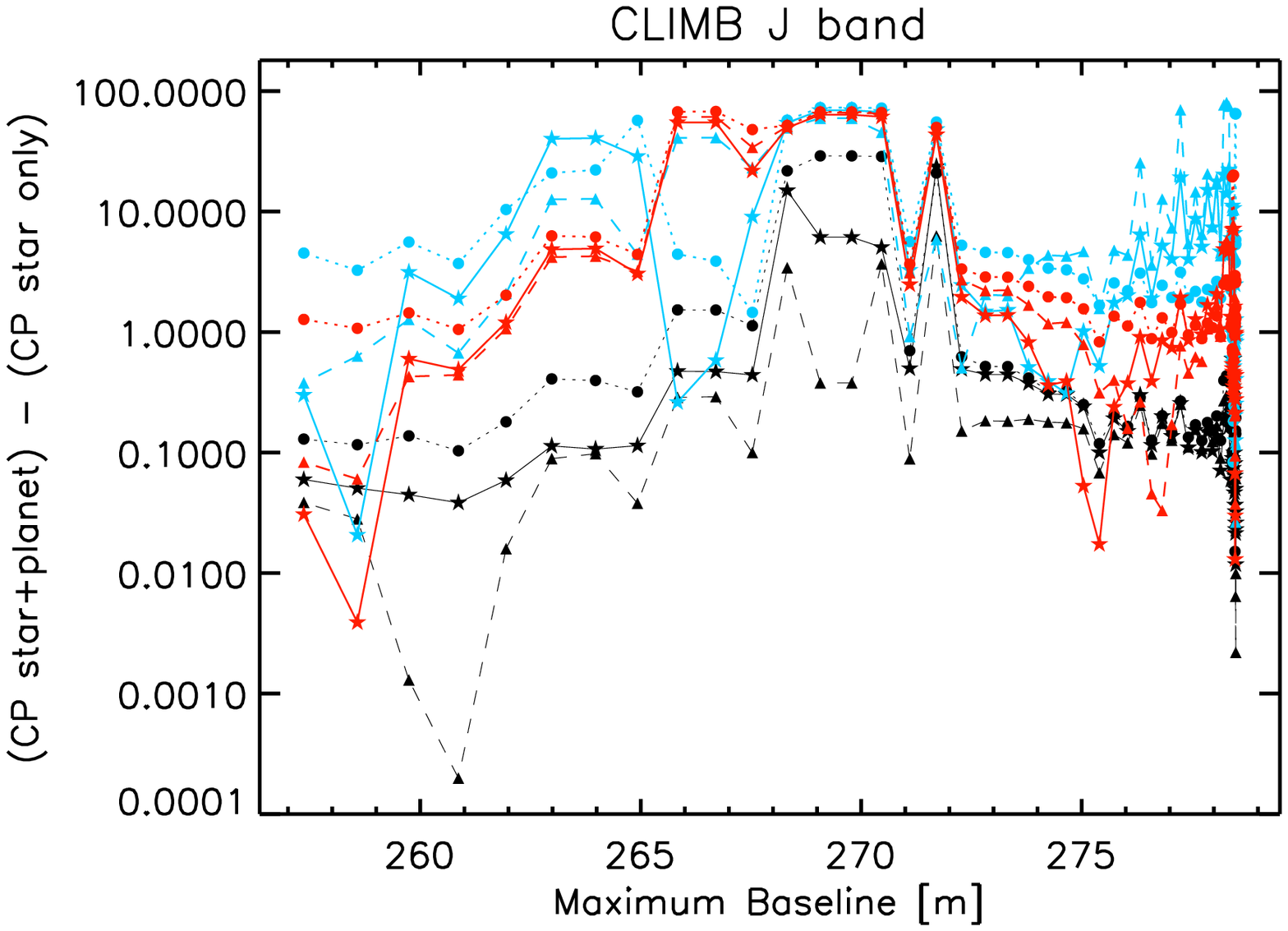}	
\includegraphics[width=0.24\hsize]{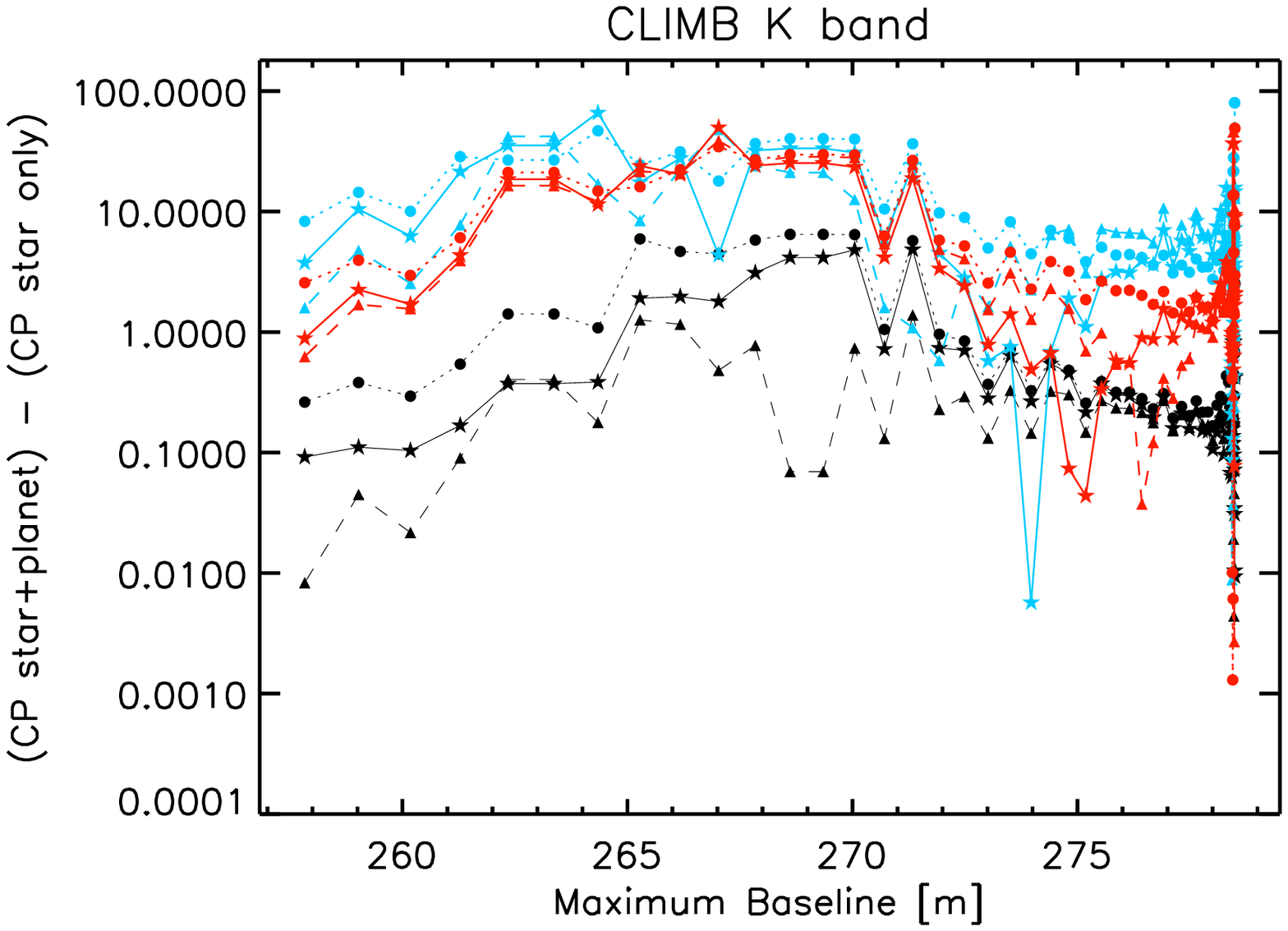}
\includegraphics[width=0.24\hsize]{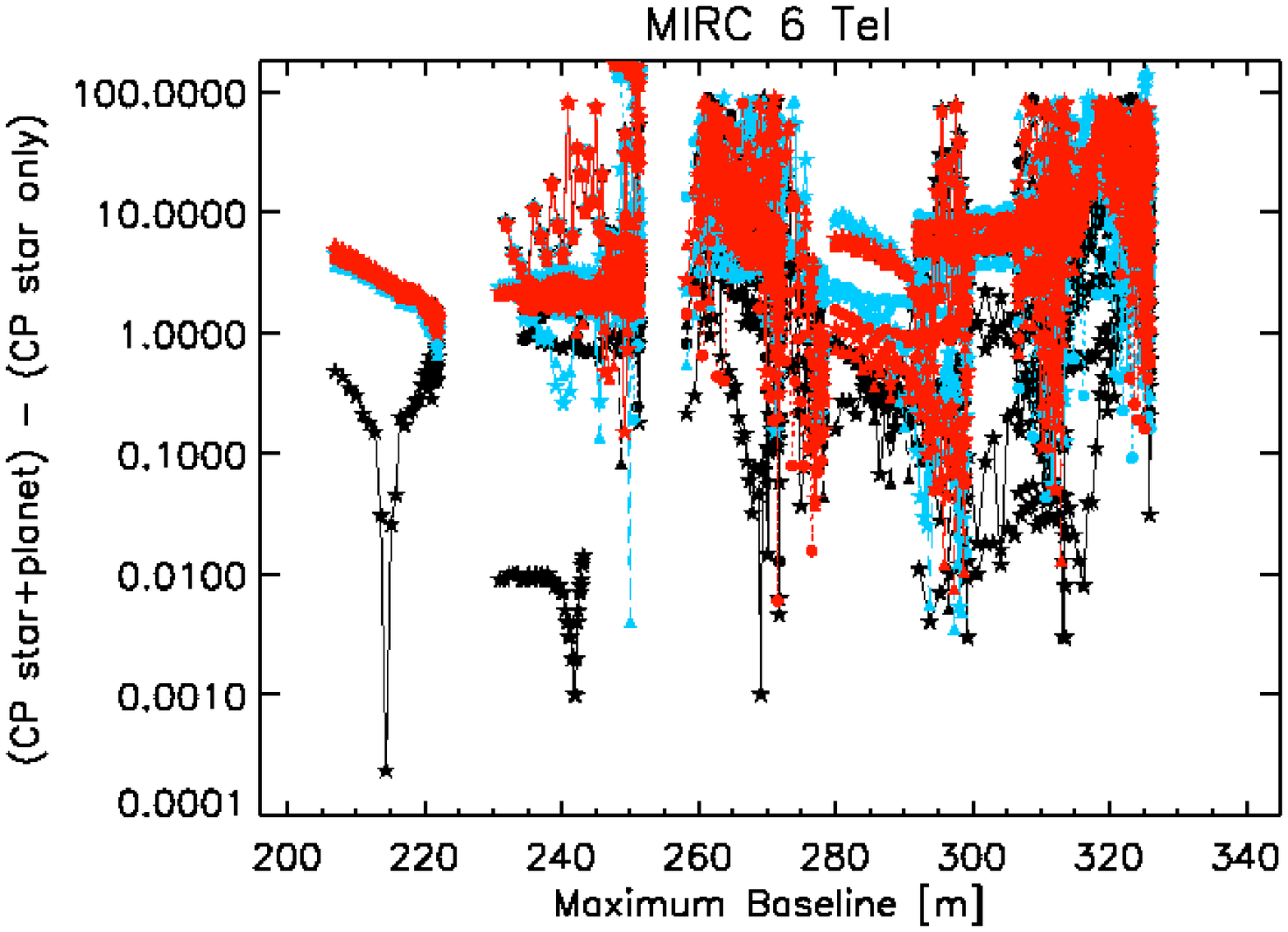}	
\includegraphics[width=0.24\hsize]{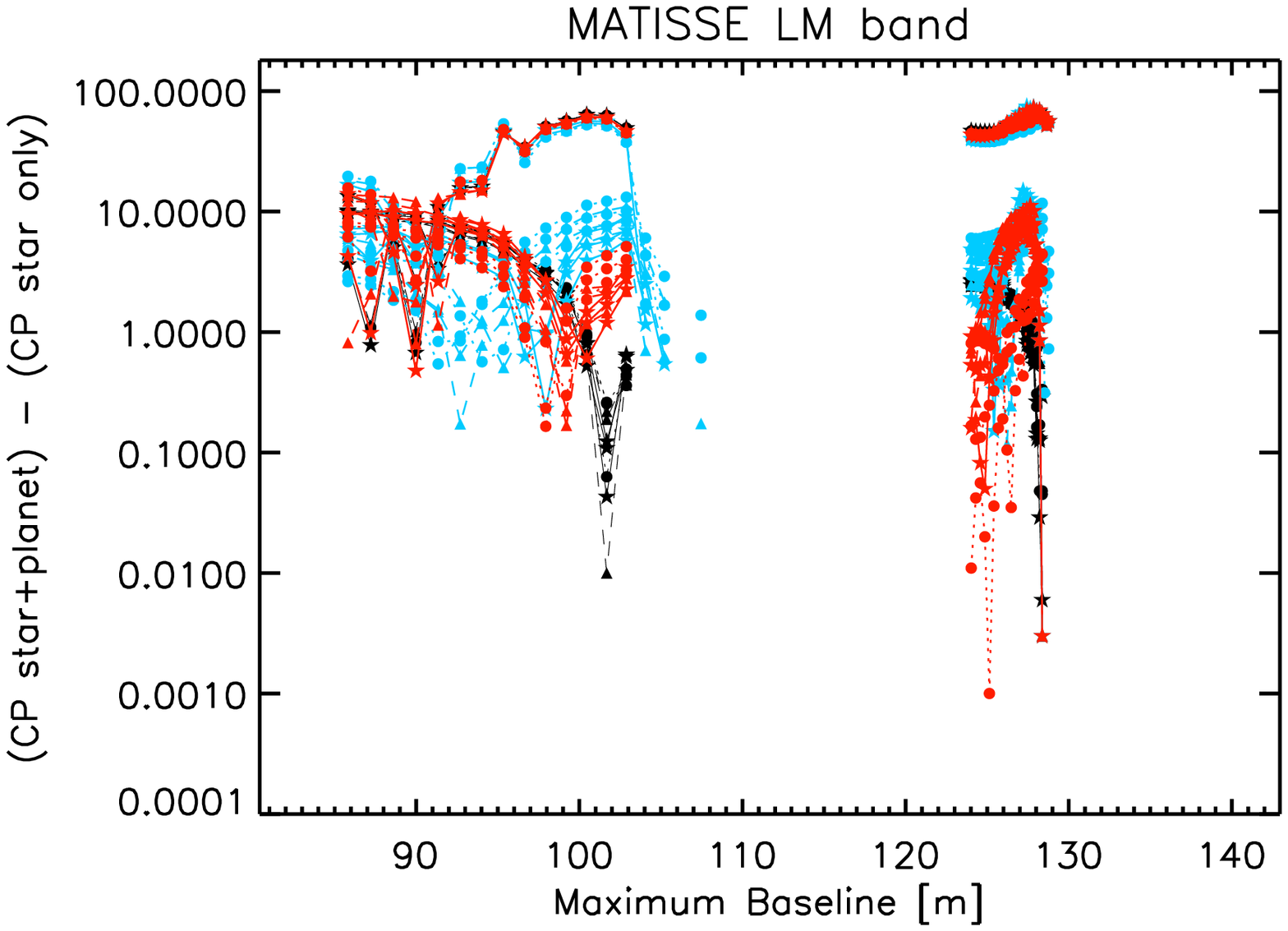}
        \end{tabular}
      \caption{Absolute closure phase differences (in degrees) between the star with a transiting planet (Fig.~\ref{transit}) and the star alone (Fig.~\ref{images}) for all the instruments of Table~\ref{instruments}. The black colour correspond to the smallest prototype planet Kepler 11-f of Table~\ref{planets}, the red to the intermediate planet HD 149026 b, and the blue to largest planet CoRoT 14-b. The star symbols connected with solid lines correspond to the planet phase entering in the stellar disk (see Fig.~\ref{transit}), the circle symbols connected with dotted line to the planet at the centre of the stellar disk, and the triangles connected with dashed line to the planet exiting the stellar disk.}
        \label{transitclosure}
   \end{figure*}

\cite{2012A&A...540A...5C} explored the impact of the convection on interferometric planetary signature around a RHD simulation of a sub-giant star. The authors estimated the impact of the granulation noise on a hot Jupiter detection using closure phases and found that there is a non-negligible and detectable contamination to the signal of the hot Jupiter due to the granulation from spatial frequencies longward of the third lobe. In this work, we extended this analysis to all the simulations of Table~\ref{simus} using the following procedure:

\begin{itemize}
\item We chose three prototypes of planets representing different sizes and compositions (Table~\ref{planets}). However, the purpose of this work is not to reproduce the exact conditions of the planet-star system detected but to employ a statistical approach on the interferometric  signature for different stellar parameters hosting planets with different sizes.
\item we simulated the transit of those planets for stars with stellar parameters of RHD simulations of Table~\ref{simus}. Two representative examples are reported in Fig.~\ref{transit};
\item we computed the closure phases for three planet-star system images corresponding to three particular planet transit phases. The different selections Instrument+Wavelength (configurations reported in Sec.~\ref{closuresect}) of the synthetic images is a representative choice among the numerous possibilities;
\item we determined the difference between the planet-star system and the star alone (Fig.~\ref{transitclosure}).
\end{itemize}
 
     \begin{table}
 \small
\centering
%\begin{minipage}[t]{\textwidth}
\caption{Prototypes of planets chosen to represent the planet transit phases of Fig.~\ref{transit}. }             % title of Table
\label{planets}      % is used to refer this table in the text
\centering                          % used for centreing table
\renewcommand{\footnoterule}{} 
\begin{tabular}{c c c c c }        % centreed columns (4 columns)
\hline\hline                 % inserts double horizontal lines
Name  & Jupiter  & Jupiter  & semi-axis & Real \\
	    & Mass   & Radius & [AU]  & hosting star  \\
	    &    &  &   & $T_{\rm{eff}}$[K]/$\log g$ \\
	    \hline
Kepler-11 f\tablefoottext{a}  &  0.006 & 0.222 & 0.2504 & 5663/4.37 \\
HD 149026 b\tablefoottext{b} & 0.360 & 0.654 & 0.0431 & 6160/4.28  \\
CoRoT-14 3b\tablefoottext{c} & 7.570 & 1.090 & 0.0269 & 6040/4.45 \\

\hline\hline                          % inserts single horizontal line
\end{tabular}
\tablefoottext{a}{\cite{2013arXiv1311.0329L}}	
\tablefoottext{b}{\cite{2005ApJ...633..465S}}
\tablefoottext{c}{\cite{2011A&A...528A..97T}} 
%\end{minipage}
\end{table}

  \begin{table*}
\tiny
\centering
%\begin{minipage}[t]{\textwidth}
\caption{The stellar intensity, $I_{\rm{star}}$, at its centre ($\mu=1$) for the synthetic images of the simulation of Fig.~\ref{transit}, and the planet integrated intensity, $I_{\rm{planet}}$, at the wavelength corresponding to the instruments of Table~\ref{instruments}. The different selections of Instrument and wavelength ($\lambda$) of the synthetic images is a representative choice among the numerous possibilities.}             % title of Table
\label{fluxratio}      % is used to refer this table in the text
\centering                          % used for centreing table
\renewcommand{\footnoterule}{} 
\begin{tabular}{l | c c c c c c }        % centreed columns (4 columns)
\hline\hline                 % inserts double horizontal lines
  Instrument         & VEGA & PAVO & NPOI  & NPOI  & CLIMB & MIRC  \\
           &              &            & 4 Tel &   6 Tel &  J band & 6 Tel \\
$\lambda$ used & 7312\AA & 6400\AA &5669\AA&5817\AA& 12862\AA & 15940\AA \\
	    \hline
	    
	     $ I_{\rm{star}}/I_{\rm{planet}}$ 	  & 1702 & 5524  &13228 & 82619 & 148 & 29 \\
%$I_{\rm{K-giant}}/I_{\rm{planet}}$	  & 603 & 1830 & 2459 & 13602 & 87 &  18\\
\hline
\hline
  Instrument       & AMBER & PIONIER & GRAVITY & PIONIER  & CLIMB & MATISSE  \\
         &                & H band &                     & K band      & K band & LM band \\
  $\lambda$ used       &23015\AA&16810\AA&22000 \AA &20510 \AA & 21350\AA & 28675\AA\\
                  \hline
$ I_{\rm{star}}/I_{\rm{planet}}$ & 84& 15& 16& 22 & 15& 13 \\
%$I_{\rm{K-giant}}/I_{\rm{planet}}$ & 62 & 10 & 12 & 15 & 9 & 10 \\

\hline\hline                          % inserts single horizontal line
\end{tabular}

%\end{minipage}
\end{table*}
 
For modeling the flux of the irradiated planet, we use the same prescriptions as \cite{2012A&A...540A...5C}. Since our main interest is related to the impact of the planet size on the interferometric observables, and the flux of the planet is much smaller than the stellar flux, we used the same model for the planetary flux \citep{2001ApJ...556..885B}, in particular the spectra of hot extrasolar planet around a generic cool star. 

Our interest is related to the closure phase signature due to the planet with respect to the stellar granulation. Figure~\ref{transit} displays the geometrical configurations of the planet-star system for the representative example of the Sun. As already reported in \cite{2012A&A...540A...5C}, the ratio between the stellar intensity and the planet integrated intensity is stronger in the infrared with respect to the optical. The stellar intensity, $I_{\rm{star}}$, at its centre ($\mu=1$) for the synthetic images of the simulations of Fig.~\ref{transit}, and the planet integrated intensity, $I_{\rm{planet}}$, at the wavelength corresponding to the instruments of Table~\ref{instruments} are reported in Table~\ref{fluxratio}.
   
   We considered three particular planet transition phases (Fig.~\ref{transit}) corresponding to the ingress and egress of the transition as well as the planet at the centre of the stellar disk. The resulting absolute differences (in degrees) between the closure phases of the planet-star system with the ones of the star alone are in Fig.~\ref{transitclosure}. The figure shows that, for all the instruments, the absolute difference scales with the size of the planet considered: the smaller planet returns smaller differences. There is however an exception for the H band-MIRC and MATISSE where there is not a clear distinction between the different planets as the baselines probe very high spatial frequencies (Fig.~\ref{uvplan}) and thus finer details.\\  
   Moreover, the closure phase differences are larger in the optical wavelengths where the stellar surface is not flat but rather ``corrugated'', due to the larger fluctuations and the higher contrast of granulation than in the infrared. Finally, while for some instruments it is not possible to disentangle the transition phase of the planet (because of the configurations chosen and/or the spatial frequencies spanned), for others (VEGA, PAVO, CLIMB, and AMBER) it is clear that different transit positions have different effects on the closure phases (this is also shown in Ligi $\&$ Mourard, in preparation).
  
  The signature of the transiting planet on the closure phase is mixed with the signal due to the convection-related surface structures. The time-scale of granulation depends on the stellar parameters, and varies from minutes or tens of minutes for solar type stars and sub-giants, to hours for more evolved red giant stars. If the transit is longer that the granulation time-scale (which is the case for most of main sequence stars), it is possible to disentangle its signal from convection by observing at particular wavelengths (either in the infrared or in the optical) and measuring the closure phases for the star at difference phases of the planetary transit. 
  
  For this purpose, it is very important to have a comprehensive knowledge of the host star to detect and characterize the orbiting planet, and RHD simulations are very important to reach this aim. 
   
    \subsection{Closure phases impact: granulation versus limb darkened law}      
   
We show in this section that the planet detection with closure phases is strongly influenced by the intrinsic stellar granulation presented in Sec.~\ref{closuresect}. For this purpose, we computed images without stellar granulation and using the limb darkened law and coefficients of \cite{2000A&A...363.1081C}. We proceeded using appropriate limb darkened coefficients for the wavelength range of the interferometric instruments of Table~\ref{instruments} and for the same stellar parameters of RHD simulations Table~\ref{simus}. Then, we simulated the planet transitions and compute the resulting closure phases using the same approach and prototypes of planets as the previous section. 
   
   \begin{figure}
   \centering
   \begin{tabular}{cc}
              \includegraphics[width=0.9\hsize]{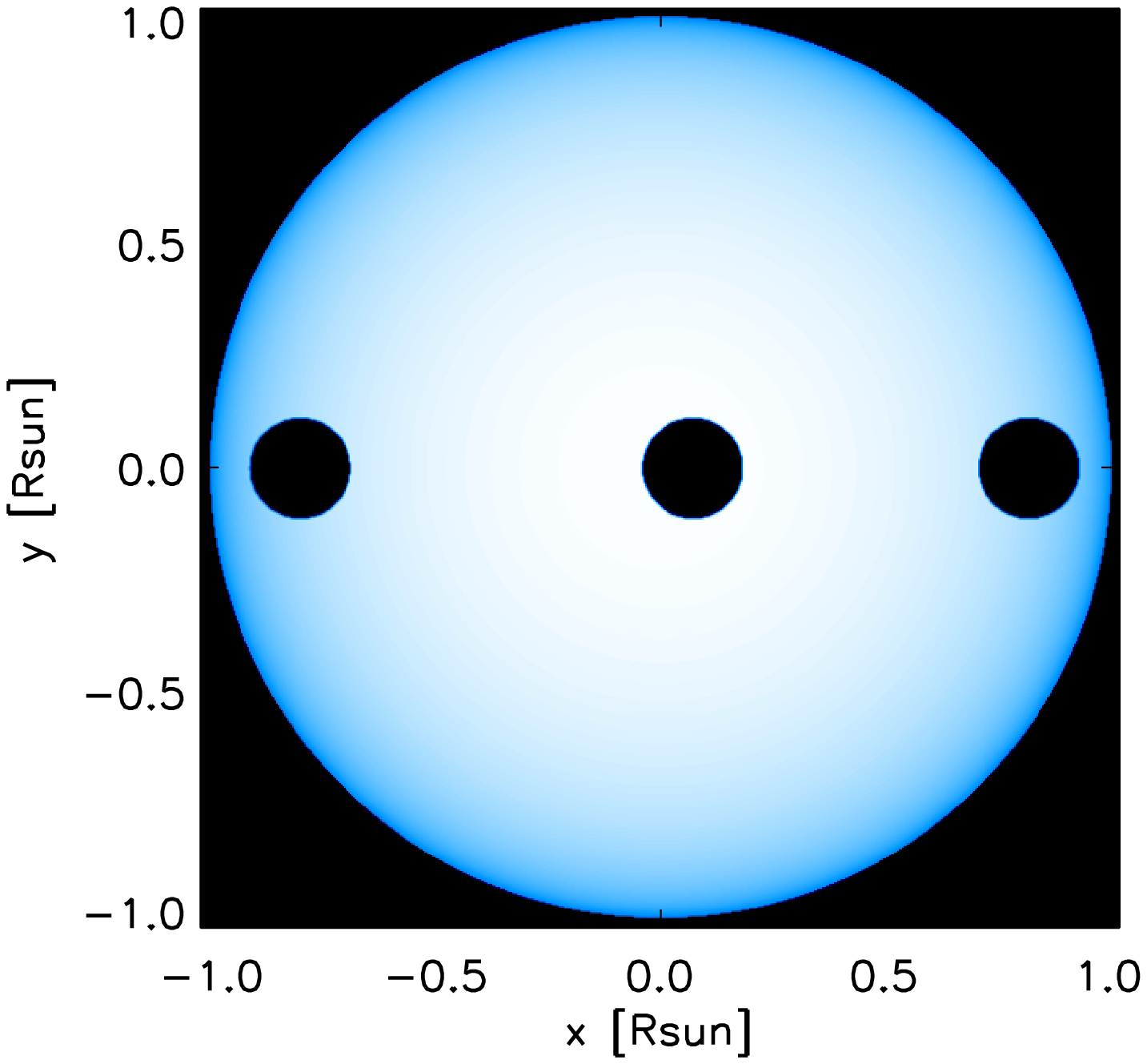}\\
              \includegraphics[width=0.9\hsize]{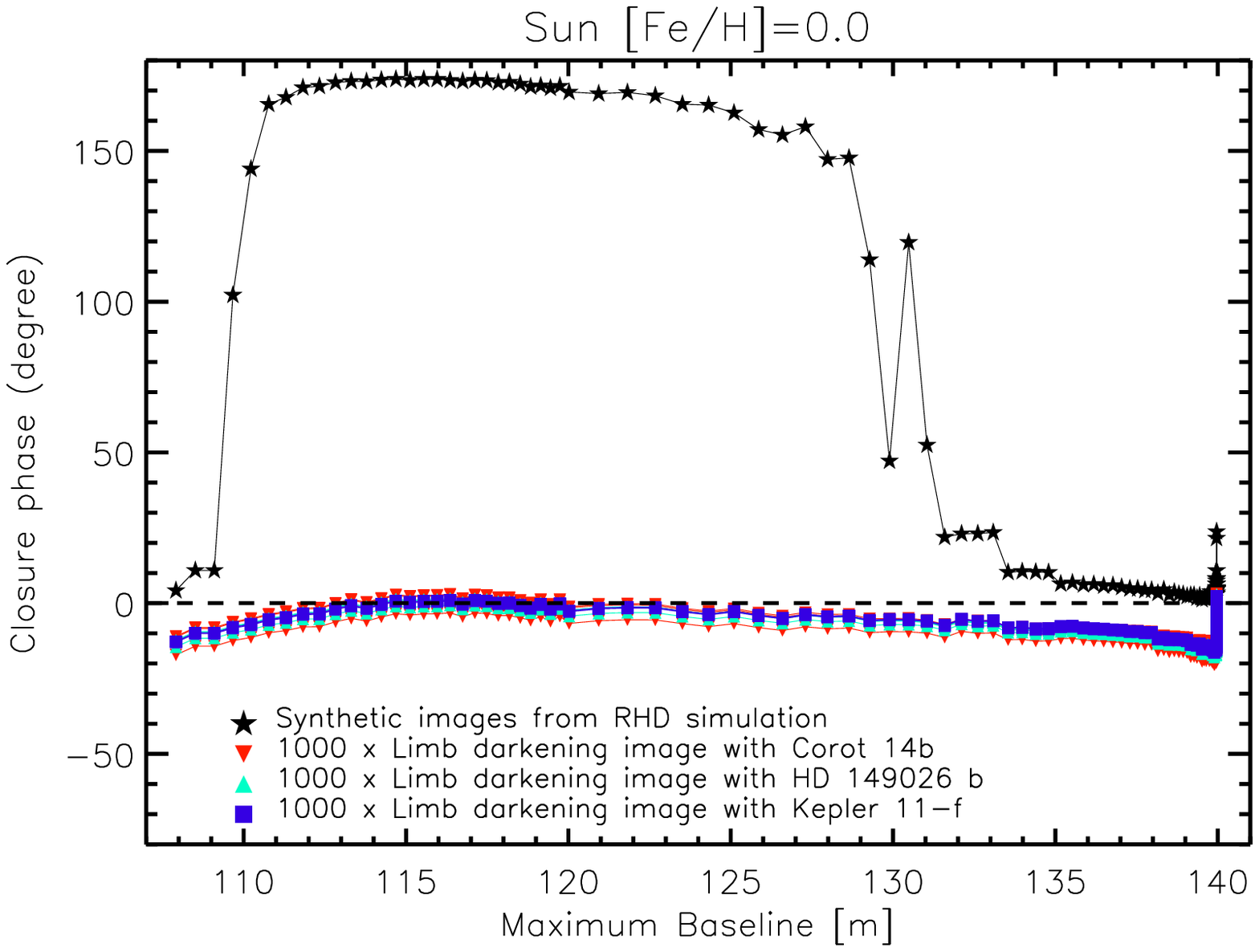}
        \end{tabular}
      \caption{\emph{Top: }limb darkened image \citep[made using the law and coefficients of][]{2000A&A...363.1081C} for AMBER instrument together with three planet transiting phases (black) for a star with the sun stellar parameters.  \emph{Bottom: } Scatter plot of closure phases computed for the Sun with transiting planets using limb-darkening unidimensional models without granulation (colored symbols) versus closure phases of the corresponding RHD simulation of Table~\ref{simus} (black symbols). The dashed line indicates the zero degree.}
        \label{LDplanet}
   \end{figure}
   
  Fig.~\ref{LDplanet} (top) displays a typical example for the limb darkened image of the Sun. The bottom panel of the figure shows the comparison between the closure phases of the synthetic images from the RHD simulation of the Sun and the ones from the limb darkened image with the transiting planet. The closure phases of a limb disk without the presence of inhomogeneities on its surface is zero or $\pm\pi$, while the transiting planet causes very small departures from spherical symmetry. It is evident that the departure from zero or $\pm\pi$ due to the convection-related surface structures are much larger than what it is expected by the transiting planet on axisymmetric images. This results is similar for all instruments and stellar parameters employed in this work. \\
It is essential to use reliable RHD hydrodynamical simulation for preparing and interpreting observations aimed to detect and characterize planets.
       
\begin{figure*}
   \centering
   \begin{tabular}{ccc}
              \includegraphics[width=0.40\hsize]{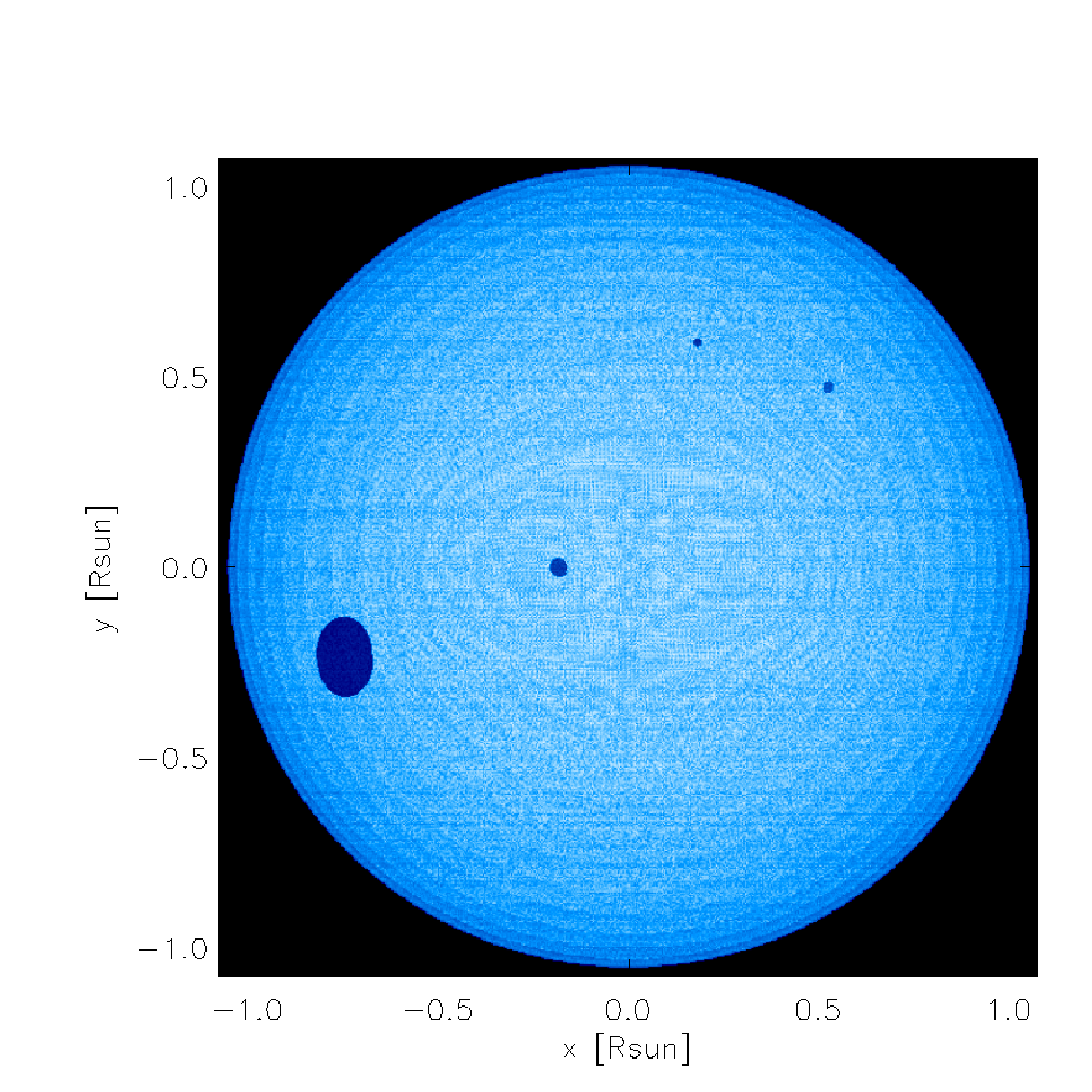}\\
              \includegraphics[width=0.40\hsize]{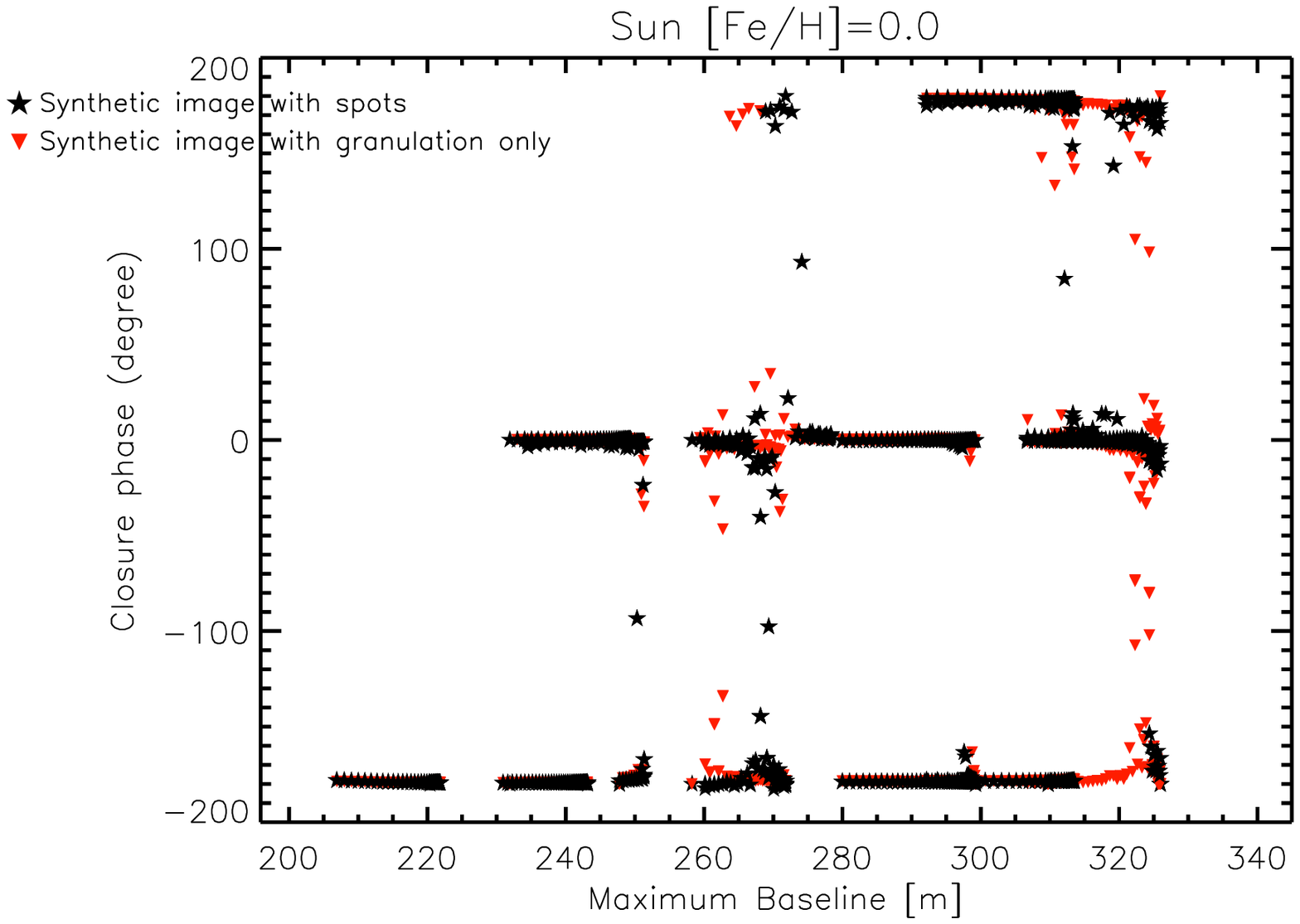}
              \includegraphics[width=0.40\hsize]{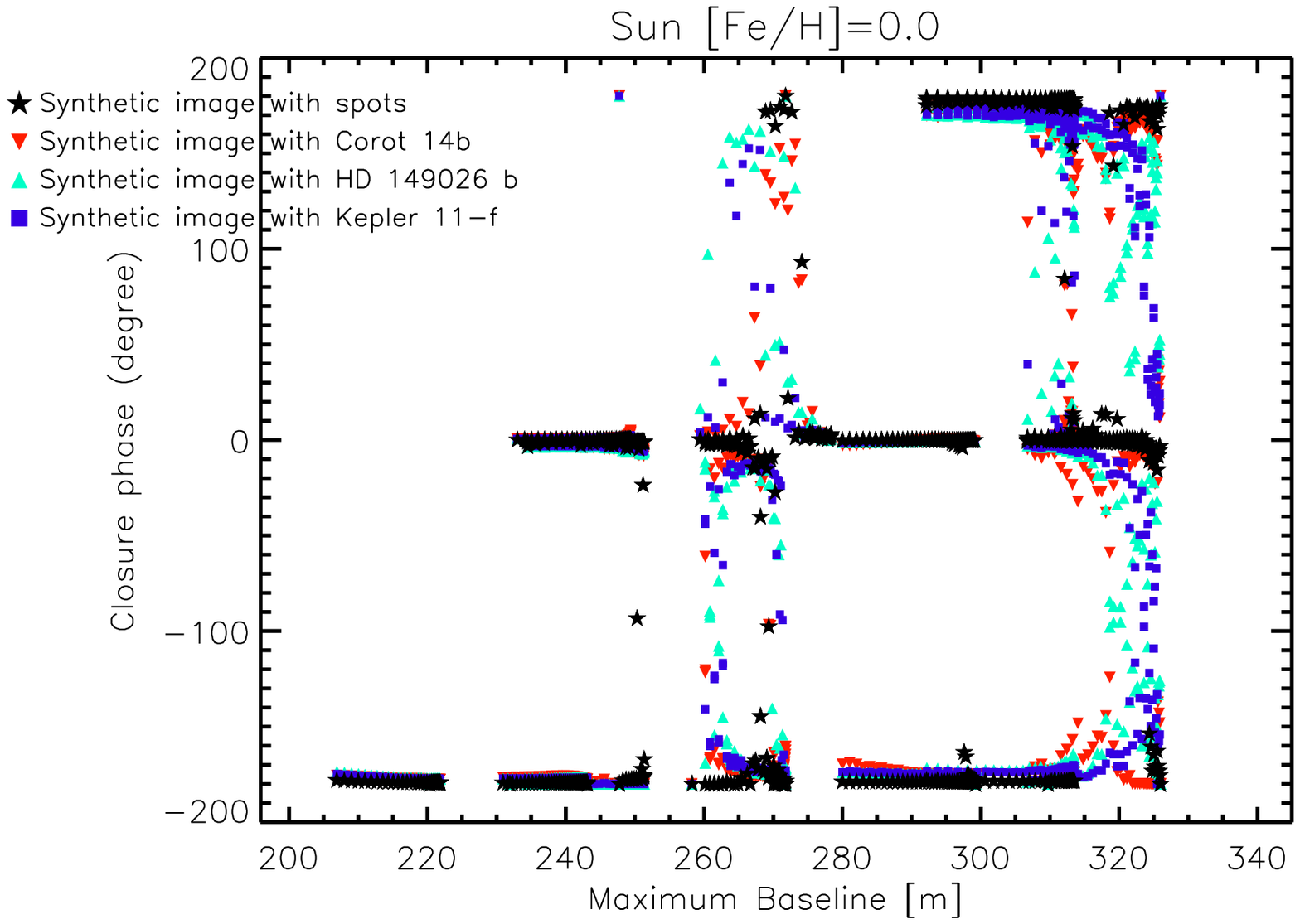}
        \end{tabular}
      \caption{\emph{Top: }synthetic stellar disk image of the Sun (Table~\ref{simus}) for MIRC instrument with four darker starspots (see text) with parameters reported in Table~\ref{starspots}. \emph{Bottom left: } Scatter plot of closure phases computed for the Sun with starspots (black stars) and for the Sun (red triangles). \emph{Bottom right: } same as in bottom left panel but for the Sun with transiting planets.}
        \label{magnetic}
   \end{figure*}             
   
\subsection{Magnetic starspots impact on closure phases}         
       
   An immediate problem for detecting transiting planets is signal contamination from starspots caused by the magnetic field of the star. Starspots are created by local magnetic field on the stellar surface and they appear as cool (and therefore dark) regions as compared to the surrounding surface. This is due to the inhibition of the convective motions by a strong enough magnetic field that blocks or redirect the energy flow from the stellar interior \citep{2009A&ARv..17..251S}. 
   We used the intensity map from the RHD simulation of the Sun (Table~\ref{simus}) in the MIRC instrument wavelengths (Table~\ref{instruments}) to construct a hypothetical starspots image. We chose to put four spots at different longitudes and distances from the center (Table~\ref{starspots}). The difference between the photosphere and the spot temperatures is up to 2000K for F and early G stars and down to 200K for late M stars \citep{2005LRSP....2....8B}, in our case we assumed temperatures for the spots of $<$2000K. We used spot size values between $\sim0.1\%$ and $\sim10\%$ of the stellar radius based on the large compilation of detected stellar spots with Doppler imaging \citep{2009A&ARv..17..251S}.\\

 \begin{table}
 \small
\centering
%\begin{minipage}[t]{\textwidth}
\caption{Parameters for the starspots.}             % title of Table
\label{starspots}      % is used to refer this table in the text
\centering                          % used for centreing table
\renewcommand{\footnoterule}{} 
\begin{tabular}{c c c c c }        % centreed columns (4 columns)
\hline\hline                 % inserts double horizontal lines
 	 & Spot 1  & Spot 2  & Spot 3 & Spot 4 \\
	    \hline
Size [$\%$ of stellar radius] &  10.0 & 1.1 & 1.5 & 2.2 \\
Temperature [K] & 3800 & 3900 & 4100 & 3900 \\
Longitude [$^\circ$] & -45. & 10. & 30. & -10. \\
\hline\hline                          % inserts single horizontal line
\end{tabular}
%\end{minipage}
\end{table}

  Figure~\ref{magnetic} (top) displays the resulting stellar disk image. The apparent size of the spots should be compared to the apparent sizes of the transiting planets of Fig.~\ref{transit}. The closure phase signal for the RHD simulation considering only the granulation and the one with starspots show non negligible differences (bottom-left panel of the figure), even thought it seems difficult to disentangle from the granulation signal due to its chaotic behavior. Moreover, it is also visible in the bottom-right panel that the starspot signal on closure phases can be of the same order as the transiting planet signal. Consequently, the  planet signal  may be contaminated. \\
Starspots caused by the magnetic field may pollute the granulation and the transiting planet signals, at least for the starspots configuration we considered. However, it should be possible to differentiate the transiting planet signal as the time-scale of a planet crossing stellar disk is much smaller than the typical rotational modulation of the star. A more detailed analysis will be reported in a forthcoming paper.
            
   \section{Conclusions}
   
   We presented an application of the \textsc{Stagger}-grid of realistic, state-of-the-art, time-dependent, radiative-hydrodynamic stellar atmosphere. We used the simulations to provide synthetic images from the optical to the infrared and extract interferometric observables aimed to study stellar convection as well as its impact on planet detection and characterisation. RHD simulations are essential for a proper quantitative analysis of interferometric observations and crucial for the extraction of the signal.
   
   We analysed the impact of convection at different wavelengths using the closure phases. Closure phase is the interferometric observable with intrinsic and unaltered information about the stellar surface asymmetries in the brightness distribution, either due to convection-related structures or a faint companion. We made our predictions as real as possible using actual interferometric instruments and configurations. All the simulations show departure from the axisymmetric case (closure phases not equal to 0 or $\pm\pi$) for all the wavelengths, but, at least for the chosen configurations, it is difficult to determine clear differences among the stellar parameters and, in particular, for the different metallicities of the solar simulations.  The levels of asymmetry and inhomogeneity of stellar disk images reach very high values of several tens of degrees with stronger effects from 3rd visibility lobe on. 
We explored the possibility of detecting the granulation pattern on two real targets (Beta Com and Procyon). We found that the detection on the 2nd lobe is possible either in the visible or in the near infrared with closure phase departures of less than 1$^\circ$; detections on the 3rd, 4th, 5th, 6th lobes (with departures up to 16$^\circ$) are possible using CHARA's instruments, and, in particular, MIRC is the most appropriate instrument because it combines good UV coverage and long baselines.
In general, interferometers probing optical and near infrared wavelengths are more adapted to reach higher spatial frequencies as the 3rd visibility lobe can be probed with baseline lengths  less than 400 meters for stellar sizes larger than 2 mas. It is more complicated for the mid-infrared wavelengths where the baselines become kilometric. We emphasise that stars should be observed at high spatial frequencies by accumulating observations on closure phases at short and long baselines. 

  We explored the impact of convection on interferometric planet signature for three prototypes of planets with sizes corresponding to one hot Jupiter, one hot Neptune, and a terrestrial one. Considering three particular planet transition phases, we compared the closure phases of the star with the transiting planet and the star alone. The signature of the transiting planet on the closure phase is mixed with the signal due to the convection-related surface structure but it is possible to disentangle it at particular wavelength (either in the infrared or in the optical).  It can be achieved by measuring the closure phases for the star at different phases of the transit. Starspots caused by the magnetic field of the star may masquerade as planets for interferometric observations. We showed that the starspot signal on closure phases can be of the same order as the transiting planet signal (at least in the example configuration we considered). However, it should be possible to differentiate between them because the time-scale of a planet crossing the stellar disk is much smaller than the typical rotational modulation of the star. It is, however, important to note that when probing high spatial frequencies, the signal to noise ratio of the measurements would be very low due to low fringe visibilities, greatly deteriorating the closure phase precision and affecting the instrument capability. Moreover, this would influence the capability and sensitivity of detecting the signatures of granulation and disentangling the planetary signal.
   
The detection and characterisation of planets must be based on a comprehensive knowledge of the host star, and this includes the detailed study of the stellar surface convection. In this context, RHD simulations are crucial to reach this aim.

\begin{acknowledgements}
   RC is the recipient of an Australian Research Council Discovery Early Career Researcher Award (project number DE120102940). This research has made use of the Exoplanet Orbit Database and the Exoplanet Data Explorer at exoplanets.org.
\end{acknowledgements}

   \bibliographystyle{aa}
\bibliography{biblio.bib}

  \end{document}